\documentclass[pre,twocolumn,twoside,superscriptaddress,floatfix]{revtex4-1}
\usepackage{natbib}

\usepackage{color}

\usepackage{graphicx,epsfig,verbatim,enumerate}
\usepackage{amssymb,amsmath,amsthm,amsbsy}
\usepackage{ifthen}
\usepackage{morefloats}

\usepackage{algorithm}
\usepackage{algpseudocode} 
\usepackage{url}
 
\usepackage{longtable}
\usepackage{booktabs}
\usepackage{tabularx}
\usepackage{longtable}
\usepackage{hyperref} 
\hypersetup{colorlinks=true, linkcolor=blue, citecolor=blue}

\usepackage{textcomp}

\usepackage{mathtools}
\usepackage{bm}

\newboolean{twocolswitch}

\newcommand{\eps}{\varepsilon}

\DeclareMathOperator{\argmax}{\arg \max}
\DeclareMathOperator{\diam}{\text{diam}}

\newcommand{\sindex}[1]{}
\newcommand{\nindex}[1]{}

\newcommand{\www}[1]{\url{#1}}

\usepackage{lettrine}

\newcommand{\dee}[1]{\textnormal{d}#1}

\newcommand{\kernel}[1]{\mathcal{K}^{(#1)}}

\begin{document}


\title{The shocklet transform: A decomposition method for the identification of local, mechanism-driven dynamics in sociotechnical time series} 


\author{\firstname{David Rushing }\surname{Dewhurst}}
\thanks{david.dewhurst@uvm.edu}
\affiliation{
  Vermont Complex Systems Center,
  Computational Story Lab,
  The University of Vermont,
  Burlington, VT 05401.
  }
\affiliation{
  Department of Mathematics \& Statistics,
  The University of Vermont,
  Burlington, VT 05401.
  }
  
\author{\firstname{Thayer }\surname{Alshaabi}}
\affiliation{
  Vermont Complex Systems Center,
  Computational Story Lab,
  The University of Vermont,
  Burlington, VT 05401.
  }
\affiliation{
  Department of Computer Science,
  The University of Vermont,
  Burlington, VT 05401.
  }

\author{Dilan Kiley \thanks{dilankiley@gmail.com}}
\affiliation{
    Zillow Group,
    Seattle, WA 98101
  }

\author{\firstname{Michael V. }\surname{Arnold}}
\affiliation{ 
  Vermont Complex Systems Center,
  Computational Story Lab,
  The University of Vermont,
  Burlington, VT 05401.
  }
 \affiliation{
  Department of Mathematics \& Statistics,
  The University of Vermont,
  Burlington, VT 05401.
  }

\author{\firstname{Joshua R. }\surname{Minot}}
\affiliation{
  Vermont Complex Systems Center,
  Computational Story Lab,
  The University of Vermont,
  Burlington, VT 05401.
  }
  \affiliation{
  Department of Mathematics \& Statistics,
  The University of Vermont,
  Burlington, VT 05401.
  }

\author{\firstname{Christopher M. }\surname{Danforth}}
\affiliation{
  Vermont Complex Systems Center,
  Computational Story Lab,
  The University of Vermont,
  Burlington, VT 05401.
  }
\affiliation{
  Department of Mathematics \& Statistics,
  The University of Vermont,
  Burlington, VT 05401.
  }

\author{\firstname{Peter Sheridan }\surname{Dodds}}
\thanks{peter.dodds@uvm.edu}
\affiliation{
  Vermont Complex Systems Center,
  Computational Story Lab,
  The University of Vermont,
  Burlington, VT 05401.
  }
\affiliation{
  Department of Mathematics \& Statistics,
  The University of Vermont,
  Burlington, VT 05401.
  }

\begin{abstract} 
We introduce a qualitative, shape-based, timescale-independent time-domain transform used to extract local dynamics from sociotechnical time series---termed the Discrete Shocklet Transform (DST)---and an associated similarity search routine, the Shocklet Transform And Ranking (STAR) algorithm, that indicates time windows during which panels of time series display qualitatively-similar anomalous behavior. After distinguishing our algorithms from other methods used in anomaly detection and time series similarity search, such as the matrix profile, seasonal-hybrid ESD, and discrete wavelet transform-based procedures, we demonstrate the DST’s ability to identify mechanism-driven dynamics at a wide range of timescales and its relative insensitivity to functional parameterization. As an application, we analyze a sociotechnical data source (usage frequencies for a subset of words on Twitter) and highlight our algorithms’ utility by using them to extract both a typology of mechanistic local dynamics and a data-driven narrative of socially-important events as perceived by English-language Twitter. 
\end{abstract}

\maketitle

\section{Introduction} \label{sec:introduction}
\begin{figure}[!ht]
\centering 
\includegraphics[width=.9\columnwidth]{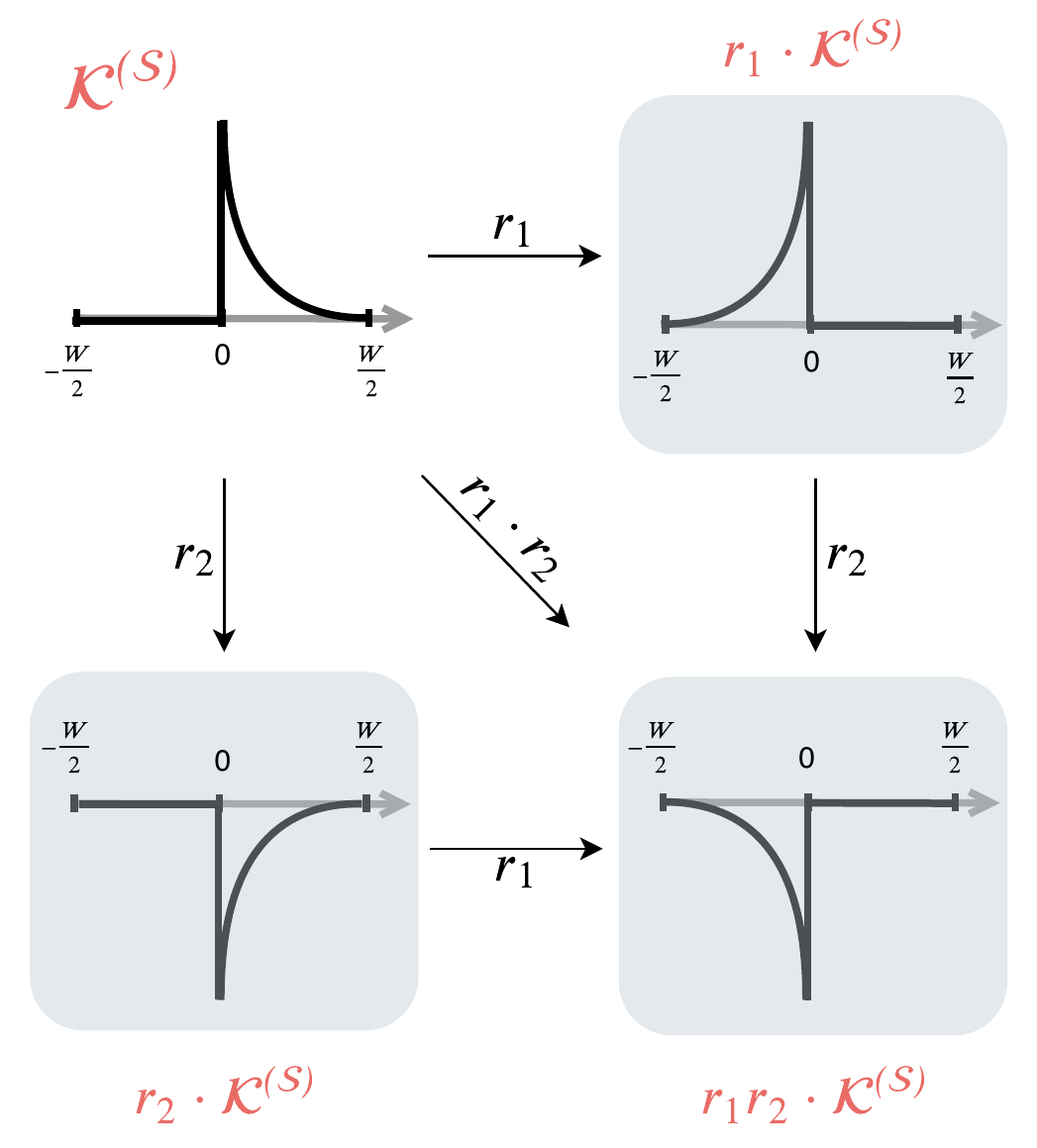}
	\caption{The discrete shocklet transform is generated through cross-correlation of pieces of shocks; this figure 
	displays effects of the action of group elements  $r_i \in R_4$ on a base ``shock-like'' kernel $\mathcal{K}$.
	The kernel $\mathcal{K}$ captures the dynamics of a constant lower level of intensity before an abrupt increase to a relatively high intensity which decays over a duration of $W/2$ units of time.
	By applying elements of $R_4$, we can effect a 
	time reversal ($r_1$) and  
	abrupt cessation of intensity followed by asymptotic 
	convergence to the prior level of intensity ($r_2$), 
	as well as the combination of these effects 
	($r_3 = r_1 \cdot r_2$).}
	\label{fig:group-action}  
\end{figure}

\begin{figure}[!ht]
\centering   
\includegraphics[width=\columnwidth]{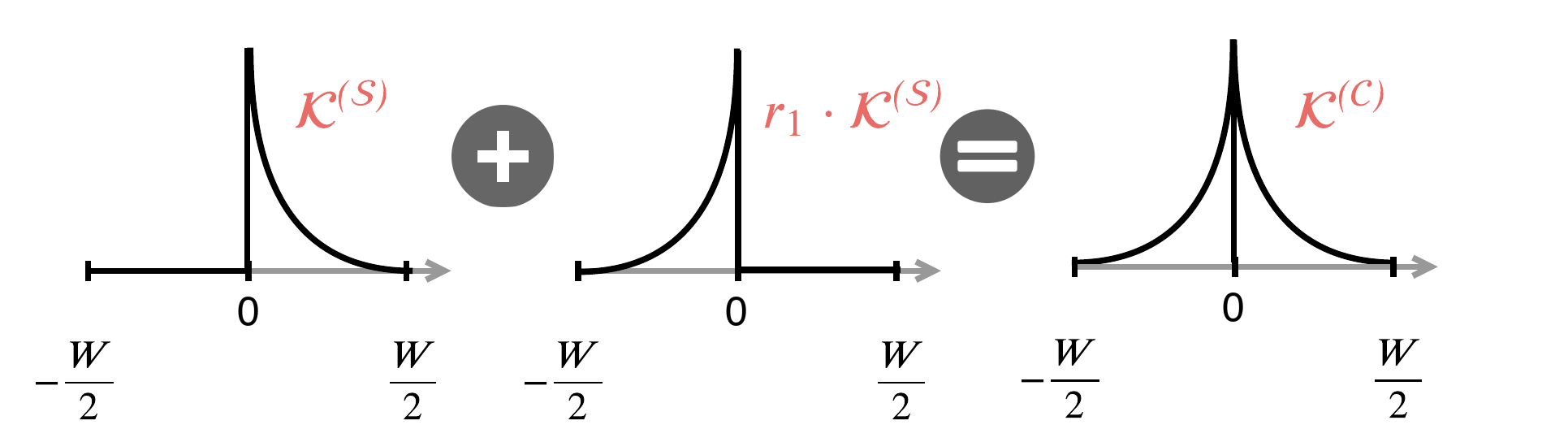}   
	\caption{This figure provides a schematic for the construction of more complicated shock dynamics from a simple initial shape ($\kernel{S}$).
	By acting on a kernel with elements $r_i$ of the reflection 
	group $R_4$ and function concatenation, 
	we create shock-like dynamics, as exemplified by the symmetric shocklet kernel 
	$ \kernel{C} = \kernel{S} \oplus [r_1 \cdot \kernel{S}]$ in this figure.
	In Section \ref{sec:typology-mech-dyn} we illuminate a typology of shock dynamics derived from combinations of these basic shapes.
	}
	\label{fig:shock-to-cusp}  
\end{figure}

\begin{figure}[!t]
    \centering 
	\includegraphics[width=\columnwidth]{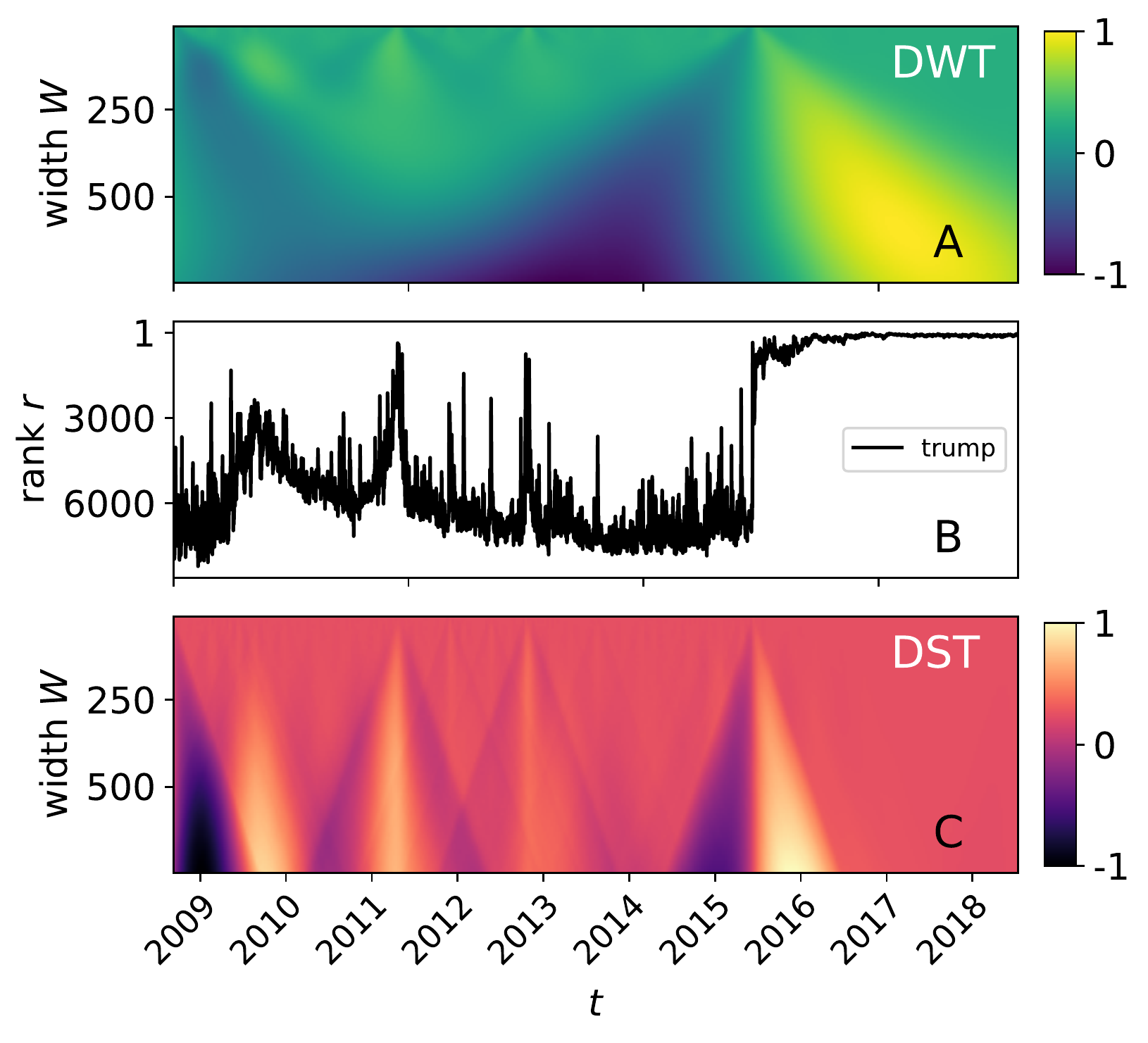}
	\caption{A comparison between the standard discrete wavelet transform (DWT) and our discrete shocklet transform 
	(DST) of a sociotechnical time series.
	Panel B displays the daily time series of the rank $r_t$ of the word ``trump'' on Twitter.
	As a comparison with the DST, we computed the DWT of $r_t$ using the 
	Ricker wavelet and display it in panel A.
	Panel C shows the DST of the time series using a symmetric power shock, 
	$\kernel{S}(\tau|W,\theta) \sim \textnormal{rect}(\tau)\tau^\theta$, with
	exponent $\theta = 3$.
	We chose to compare the DST with the DWT because the DWT is similar in mathematical construction (see Appendix \ref{app:stats} for a more extensive discussion of this assertion), but differs in the choice of convolution kernel (a wavelet, in the case of the DWT, and a piece of a shock, in the case of the DST) and the method by which the transform accounts for signal at multiple timescales. 
	} 
	\label{fig:dwt_vs_dcut} 
\end{figure}

The tasks of peak detection, similarity search, and anomaly detection
in time series is often accomplished using the discrete wavelet transform (DWT)
\cite{chaovalit2011discrete} or matrix-based methods \cite{yeh2017matrix,zhu2018introducing}.
For example, wavelet-based methods have been used for outlier detection in financial time series 
\cite{struzik2002wavelet}, similarity search and compression of various correlated time series
\cite{popivanov2002similarity},
signal detection in meteorological data \cite{lau1995climate},
and homogeneity of variance testing in time series with long memory \cite{whitcher2002testing}.
Wavelet transforms have far superior localization in the time domain than do
pure frequency-space methods such as the short-time Fourier transform \cite{benitez2010wavelet}.
Similarly, the chirplet transform is used in the analysis of 
phenomena displaying periodicity-in-perspective (linearly- or quadratically-varying frequency), 
such as images and radar signals
\cite{mann1991chirplet,wang2002moving,spanos2007time,taebi2016effect}.
Thus, when analyzing time series that are partially composed of exogenous shocks and endogenous shock-like local dynamics, we should use a small sample of such a function---a ``shock'', examples of which are depicted in
Fig.\ \ref{fig:group-action}, and functions generated by concatenation of these building blocks, such as that shown in Fig.\ \ref{fig:shock-to-cusp}. 
In this work, we introduce the Discrete Shocklet Transform (DST), generated by
cross-correlation functions
of a shocklet.
As an immediate example (and before any definitions or technical discussion),
we contrast the DWT with the DST of a sociotechnical time series---popularity of the word ``trump'' on the social media website Twitter---in Fig.\ \ref{fig:dwt_vs_dcut}, 
which is a visual display of what we claim is the DST's suitability for detection of local mechanism-driven dynamics in time series.
\medskip

\noindent
We will show that the DST can be used to extract shock and shock-like dynamics of particular interest from time series through construction of an 
indicator function that compresses time-scale-dependent information into a single spatial dimension using 
prior information on timescale and parameter importance.
Using this indicator, we are able to highlight windows in which underlying 
mechanistic dynamics are hypothesized to contribute a stronger 
component of the signal than purely stochastic dynamics, and 
demonstrate an algorithm---the Shocklet Transform and Ranking (STAR) algorithm---by which we are able to automate 
\textit{post facto} detection of endogenous, mechanism-driven dynamics. 
As a complement to techniques of changepoint analysis, methods by which one can detect changes in the level of time series 
\cite{page1955test,mallat1992singularity},
the DST and STAR algorithm detect changes in the underlying mechanistic local dynamics of the time series.
Finally, we demonstrate a potential usage of the shocklet transform by applying it to the LabMT Twitter dataset
\cite{dodds2011temporal} to extract word usage time-series matching the qualitative form of a shock-like kernel at multiple timescales.

\section{Data and Theory} \label{sec:theory}
 
\subsection{Data}\label{sec:data}
Twitter is a popular micro-blogging service that allows users to share thoughts and news with a global 
community via short messages (up to 140 or, from around November 2017 on, 280 
characters, in length).
We purchased access to Twitter's ``decahose'' streaming API and used it to collect a random 10\% sample
of all public tweets authored between September 9, 2008 and April 4, 2018
\cite{li2017much}.
We then parsed these tweets to count appearances of words included in the LabMT dataset, a set of roughly 10,000 of the most commonly used words in English \cite{dodds2011temporal}. 
The dataset has been used to construct nonparametric sentiment analysis models 
\cite{reagan2017sentiment} and forecast mental illness \cite{reece2017forecasting}
among other applications 
\cite{frank2013happiness,mitchell2013geography,lemahieu2015optimizing}. 
From these counts, we analyze the time series of word popularity as measured by rank of word usage: on day $t$, the most-used word is assigned rank 1, the second-most assigned rank 2, and so on to create time series of word rank $r_t$ for each word.
\subsection{Theory}

\subsubsection{Algorithmic details: description of the method}
There are multiple fundamentally-deterministic mechanistic models for local dynamics of sociotechnical time series. 
Nonstationary local dynamics are generally well-described by exponential, bi-exponential, or power-law decay functions;
mechanistic models thus usually generate one of these few functional forms.
For example, Wu and Huberman described a stretched-exponential model for collective human attention
\cite{wu2007novelty}, and
Candia \textit{et al.} derived a biexponential function for collective human memory on longer timescales
\cite{candia2019universal}. 
Crane and Sornette assembled a Hawkes process for video views that produces 
power-law behavior by using power-law excitement kernels \cite{crane2008robust}, and
Lorenz-Spreen \textit{et al.}\ demonstrated a speeding-up dynamic in collective social attention mechanisms \cite{lorenz2019accelerating},
while De Domenico and Altmann put forward a stochastic model incorporating social heterogeneity and 
influence \cite{de2019unraveling},
and Ierly and Kostinsky introduced a rank-based, signal-extraction method with applications to meteorology data \cite{1906.08729}.
In Sec.\ \ref{sec:comparison} we conduct a literature review, contrasting our methods with existing anomaly detection and similarity search time series data mining algorithms and demonstrating that the DST and associated STAR algorithm differ substantially from these existing algorithms.
We have open-sourced implementations of the DST and STAR algorithm; code for these implementations is available at a publicly-accessible repository \footnote{
Python implementations of the DST and STAR algorithms are located at this 
git repository: 
\href{https://gitlab.com/compstorylab/discrete-shocklet-transform}{https://gitlab.com/compstorylab/discrete-shocklet-transform}
}.
\medskip

\noindent
We do not assume any specific model in our work. 
Instead, by default we define 
a kernel $\kernel{\cdot}$ as one of a few basic functional forms: exponential growth,
\begin{equation}\label{eq:exp-kernel}
    \kernel{S}(\tau|W,\theta) \sim \textnormal{rect}(\tau - \tau_{0})e^{\theta (\tau - \tau_{0})};
\end{equation}
monomial growth, 
\begin{equation}\label{eq:power-kernel}
    \kernel{S}(\tau|W,\theta) \sim \textnormal{rect}(\tau - \tau_0)\tau^\theta;
\end{equation}
power-law decay,
\begin{equation}\label{eq:power-law-shock}
    \kernel{S}(\tau|W,\theta) \sim \textnormal{rect}(\tau - \tau_{0})
    |\tau - \tau_0 + \varepsilon|^{-\theta},
\end{equation}
or sudden level change (corresponding with a changepoint detection problem),
\begin{equation}\label{eq:spike-kernel}
      \kernel{Sp}(\tau|W,\theta) \sim \textnormal{rect}(\tau - \tau_{0})[\Theta(\tau) - \Theta(-\tau)],
\end{equation}
where $\Theta(\cdot)$ is the Heaviside step function.
The function $\textnormal{rect}$ is the rectangular function ($ \textnormal{rect}(x)=1$ for $0<x<W/2$ and $\textnormal{rect}(x) = 0$ otherwise), 
while in the case of the power-law kernel we add a constant $\varepsilon$ to ensure nonsingularity.
The parameter $W$ controls the support of $\kernel{\cdot}(\tau|W,\theta)$; the kernel is identically zero outside of the interval 
$[\tau - W/2, \tau + W/2]$.
We define the window parameter $W$ as follows: moving from a window size of $W$ to a window size of $W + \Delta W$
is equivalent to upsampling the kernel signal
by the factor $W + \Delta W$, applying an ideal lowpass filter, and downsampling by the factor $W$.
In other words, if the kernel function $\kernel{\cdot}$ is defined for each of
$W$ linearly spaced points between $-N/2$ and $N/2$, moving to a window size of $W$ to $W + \Delta W$ is equivalent to computing $\kernel{\cdot}$ for 
each of $W + \Delta W$ linearly-spaced points between $-N/2$ and $N/2$.
This holds the dynamic range of the kernel constant while accounting for the dynamics described by the kernel at all timescales of interest.
We enforce the condition that $\sum_{t=-\infty}^{\infty} \kernel{\cdot}(t| W,\theta) = 0$ for any window size $W$.
\medskip

\noindent
It is decidedly not our intent to delve into the question of how and why deterministic underlying dynamics in sociotechnical systems arise. 
However, we will provide a brief justification for the functional forms of the kernels presented in the last paragraph as scaling solutions to a variety of parsimonious models of local deterministic dynamics:
\begin{itemize}
    \item If the time series $x(t)$ exhibits exponential growth with a state-dependent growth damper $D(x)$, the dynamics can be described by 
    \begin{equation}
        \frac{\dee x(t)}{\dee t} = \frac{\lambda}{D(x(t))}x(t),\ x(0) = x_0.
    \end{equation}
    If $D(x) = x^{1/n}$, the solution to this IVP scales as $x(t) \sim t^n$, which is the functional form given in 
    Eq.\ \ref{eq:power-kernel}. When $D(x) \propto 1$ (i.e., there is no damper on growth) then the solution is an exponential function, the functional form of Eq.\ \ref{eq:exp-kernel}.
    \item If instead the underlying dynamics correspond to exponential decay with a time- and state-dependent half-life $\mathcal{T}$, we can model the dynamics by the system
    \begin{align}
        \frac{\dee x(t)}{\dee t} &= -\frac{x(t)}{\mathcal{T}(t)},\ x(0) = x_0 
        \label{eq:scaling-halflife}\\
        \frac{\dee \mathcal{T}(t)}{\dee t} &= f(\mathcal{T}(t), x(t)),\ \mathcal{T}(0) = \mathcal{T}_0.
    \end{align}
    If $f$ is particularly simple and given by $f(\mathcal{T}, x) = c$ with $c > 0$, then the 
    solution to Eq.\ \ref{eq:scaling-halflife} scales as 
    $x(t) \sim t^{-1/c}$, the functional form of Eq.\ \ref{eq:power-law-shock}.
    The limit $c \rightarrow 0^+$ is singular and results in dynamics of exponential decay, given by reversing time in Eq.\ \ref{eq:exp-kernel} (about which we expound later in this section).
    \item As another example, the dynamics could be essentially static except when a latent variable $\varphi$ changes state or moves past a threshold of some sort:
    \begin{align}
       \frac{\dee x(t)}{\dee t} &= \delta\left( 
       \varphi(t) - \varphi^*
       \right),\ x(0) = x_0\\
       \frac{\dee \varphi(t)}{\dee t} &= g(\varphi(t), x(t)),
       \varphi(0) = \varphi_0.
    \end{align}
    In this case the dynamics are given by a step function from $x_0$ to $x_0 + 1$ the first time $\varphi(t)$ changes position relative to $\varphi^*$, and so on; these are the dynamics
    we present in Eq.\ \ref{eq:spike-kernel}.
\end{itemize}
This list is obviously not exhaustive and we do not intend it to be so.
\medskip

\noindent
We can use kernel functions $\kernel{\cdot}$ as basic building blocks of richer local mechanistic dynamics through function concatenation and the operation of the two-dimensional reflection group $R_4$.
Elements of this group correspond to $r_0 = \text{id}$, $r_1 = $ reflection across the vertical axis (time reversal), $r_2 = $ negation (e.g., from an increase in usage frequency to a decrease in usage frequency), and $r_3 = r_1 \cdot r_2 = r_2 \cdot r_1$.
We can also model new dynamics by concatenating kernels, i.e., ``glueing'' kernels back-to-back. 
For example, we can generate ``cusplets'' with both anticipatory and relaxation dynamics by concatenating a shocklet $\kernel{S}$ with a time-reversed 
copy of itself:
\begin{equation}\label{eq:cusp-kernel}
	\kernel{C}(\tau|W,\theta) \sim \kernel{S}(\tau|W,\theta) \oplus [r_1 \cdot \kernel{S}(\tau|W,\theta)].
\end{equation}
We display an example of this concatenation operation in Fig.\ \ref{fig:shock-to-cusp}. For much of the remainder of the work, we conduct analysis using this symmetric kernel.
\medskip

\noindent
The discrete shocklet transform (DST) of the time series $x(t)$ is defined by  
\begin{equation}\label{eq:discrete-shocklet-transform}
\mathrm{C}_{\kernel{S}}(t, W|\theta) = 
	\displaystyle\sum_{\tau=-\infty}^{\infty} x(\tau + t)\kernel{S}(\tau|W,\theta),
\end{equation} 
which is the cross-correlation of the sequence and the kernel.
This defines a $T \times N_W$ matrix containing an entry for each point in time $t$ and window width $W$ considered.
\medskip

\begin{figure} 
\centering
	\includegraphics[width=\columnwidth]{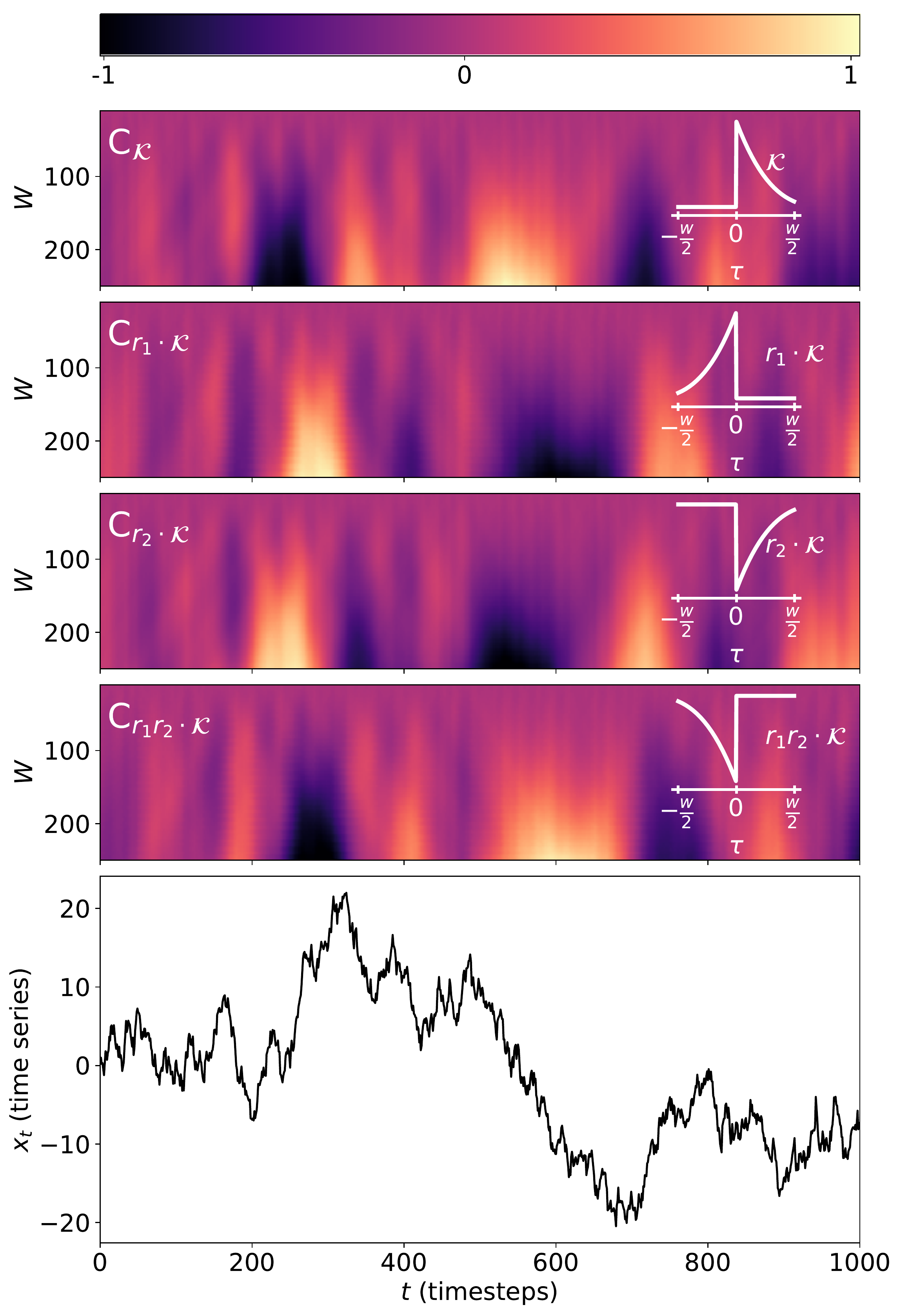}
	\caption{Effects of the reflection group $R_4$ on the shocklet transform.
	The top four panels display the results of the shocklet transform of  
	a random walk $x_t = x_{t-1} + z_t$ with $z_t \sim \mathcal{N}(0,1)$,
	displayed in the bottom panel,
	using the kernels $r_j \cdot \kernel{S}$, where $r_j \in R_4$.
	}
	\label{fig:reflection-group}
\end{figure}

\noindent
To convey a visual sense of what the DST looks like when using a shock-like, asymmetric kernel, we compute the DST of a random walk
$x_t - x_{t-1} = z_t$ (we define
$z_t \sim \mathcal{N}(0,1)$) using a kernel function 
$\kernel{S}(\tau|W, \theta) \sim \textnormal{rect}(\tau)\tau^\theta$ with $\theta = 3$
and display the resulting matrix for window sizes $W \in [10, 250]$ 
in Fig.\ \ref{fig:reflection-group}.
The effects of time reversal by action of $r_1$ are visible when comparing the first and third panels with the second and fourth panels, and the result of negating the kernel by acting on it with $r_2$ is apparent in the negation of the matrix values when comparing the first and second panels and with the third and fourth.
For this figure, we used a random walk as an example time series here as there is, by definition, no underlying generative mechanism causing any shock-like dynamics; these dynamics appear only as a result of integrated noise. 
We are equally likely to see large upward-pointing shocks as large downward-pointing shocks because of this, which allows us to see the activation of both upward-pointing and downward-pointing kernel functions.
\medskip

\begin{figure*}
\centering 
	\includegraphics[width=\textwidth]{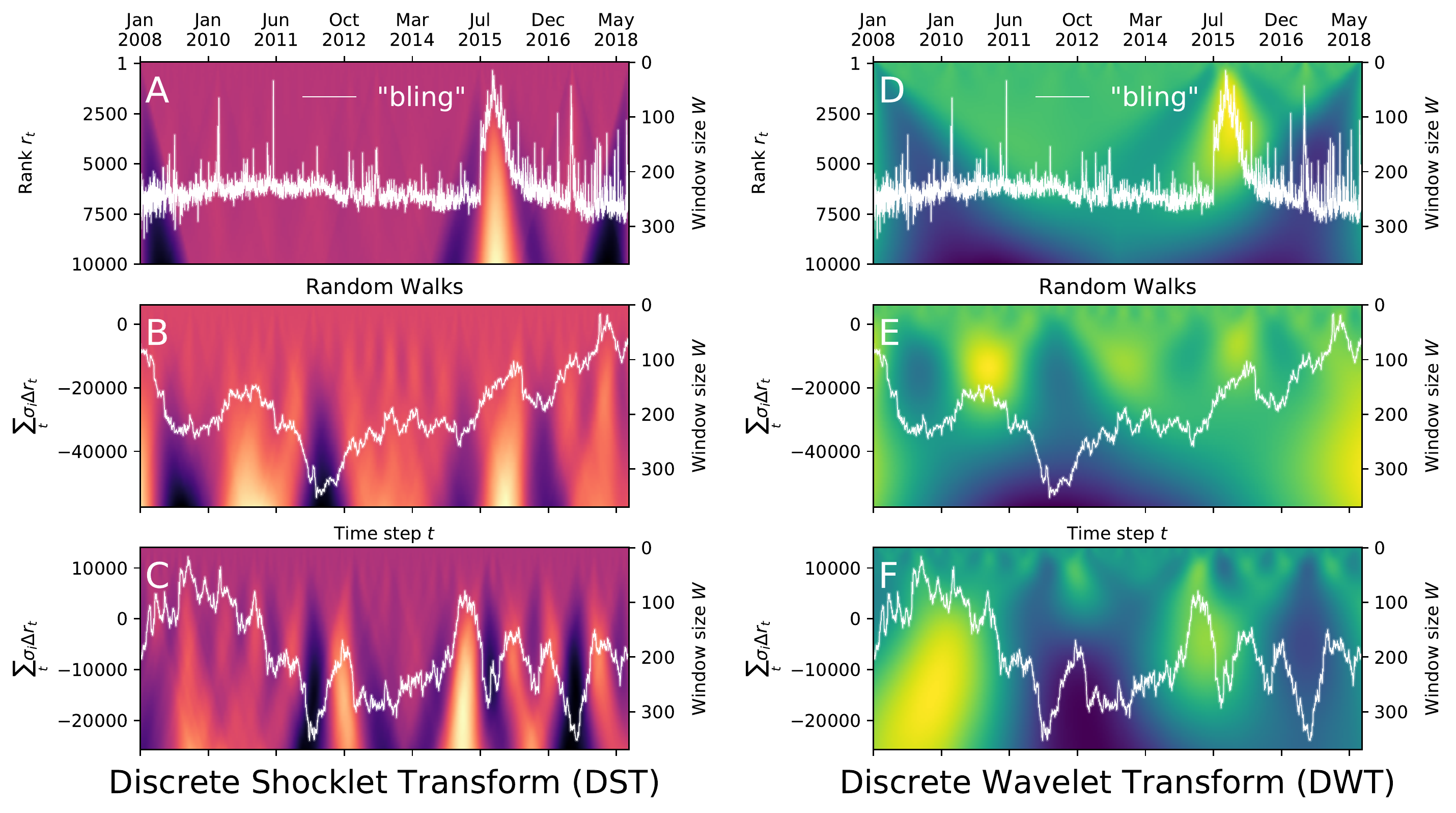}
	\caption{
	Intricate dynamics of sociotechnical time series.
	Panels A and D show the time series of the ranks down from top of the word ``bling'' on Twitter.
	Until mid-summer 2015, the time series presents as random fluctuation about a steady, relatively-constant level. 
	However, the series then displays a large fluctuation, increases rapidly, and then decays slowly after a sharp peak. 
	The underlying mechanism for these dynamics was the release of a popular song titled ``Hotline Bling''.
	To demonstrate the qualitative difference of the ``bling'' time series from draws from a null random walk model, the details of which are given in Appendix \ref{app:stats}.
	Panels A, B, and C show the discrete shocklet transform of the original series for ``bling'' 
	and the random walks $\sum_{t'\leq t} \Delta r_{\sigma_i t}$,
	showing the responsiveness of the DST to nonstationary 
	local dynamics and its insensitivity to dynamic range.
	Panels D, E, and F, on the other hand, display the discrete 
	wavelet transform of the original series and 
	of the random walks,  
	demonstrating the DWT's comparatively less-sensitive nature to local
	shock-like dynamics.
	}
	\label{fig:symmetric-group-null-example}
\end{figure*}

\noindent
As a comparison with this null example, we computed the DST of a sociotechnical time series, the rank of the word ``bling'' among the LabMT words on Twitter, and two draws from a null random walk model, and displayed the results in Fig. \ref{fig:symmetric-group-null-example}.
Here, we calculated the DST using the symmetric kernel given in Eq.\ \ref{eq:cusp-kernel}.
(For more statistical details of the null model, see Appendix \ref{app:stats}.)
We also computed the DWT of each of these time series and display the resulting wavelet transform matrices next to the shocklet transform matrices in Fig.\ \ref{fig:symmetric-group-null-example}.
Direct comparison of the sociotechnical time series ($r_t$) with the draws from the null models reveals $r_t$'s moderate autocovariance as well as the large, shock-like fluctuation that occurs in late July of 2015. (This underlying driver of this fluctuation was the release of a popular song entitled ``Hotline Bling'' on July 31st, 2015.)
In comparison, the draws from the null model have a covariance with much more prominent time scaling and do not exhibit dramatic shock-like fluctuations as does $r_t$.
Comparing the DWT of these time series with the respective DST provides more evidence that the DST exhibits superior space-time localization of shock-like dynamics than does the DWT.
\medskip

\begin{figure*}[!ht]
\centering
	\includegraphics[width=\textwidth]{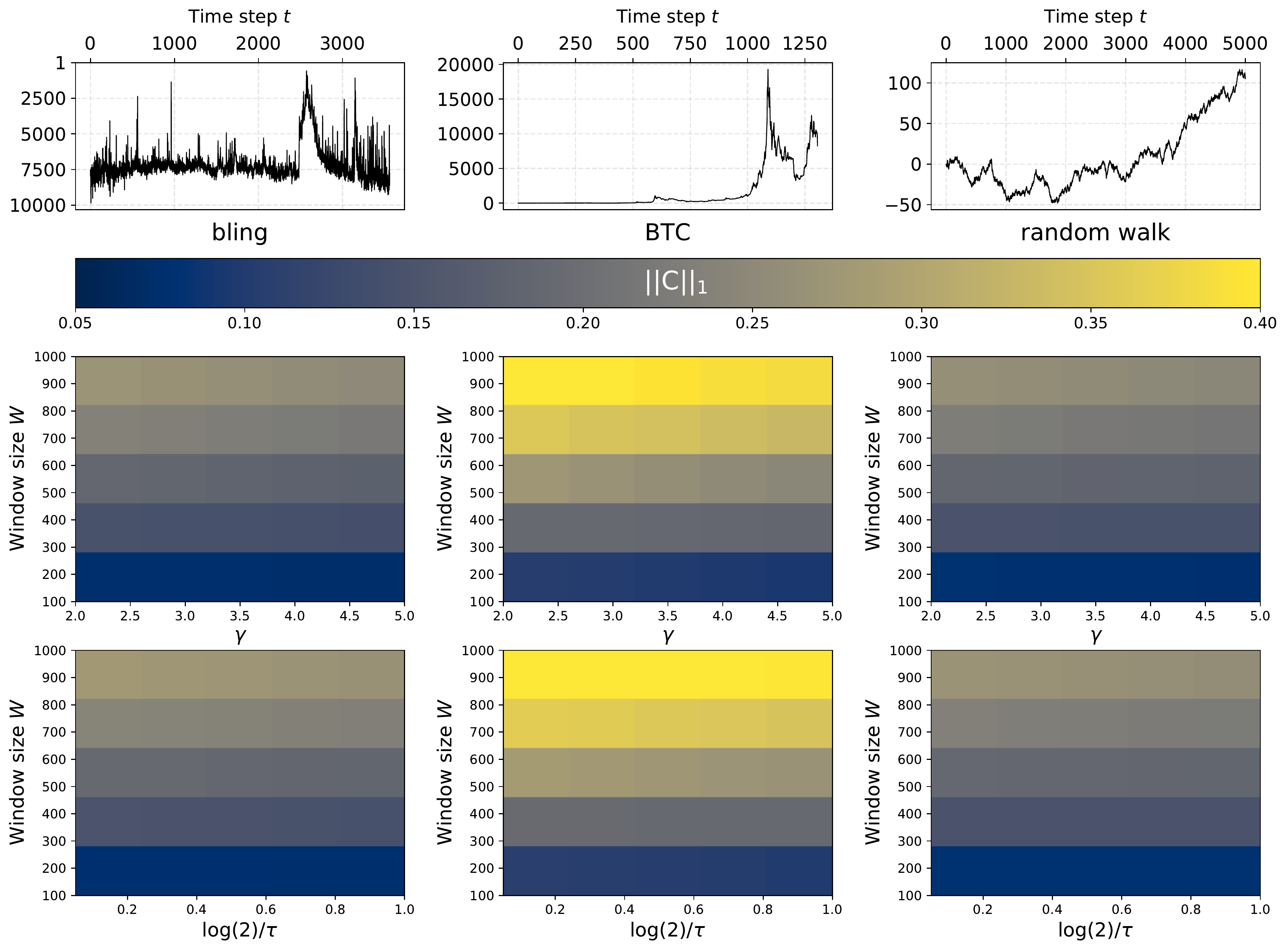}
	\caption{
    The shock indicator function is relatively insensitive to functional forms $\kernel{\cdot}$ and values of the kernel's parameter vector $\theta$ so long as the kernel functions are qualitatively similar (e.g., for cusp-like dynamics---as considered in this figure and in Eq.\ \ref{eq:cusp-kernel}---$\kernel{C}$ displaying increasing rates of increase followed by decreasing rates of decrease).
    Here we have computed the shock indicator function $\mathrm{C}_{\kernel{S}}(\tau|\theta)$ (Eq.\ \ref{eq:sif}) for three different time series: two sociotechnical and one null example. 
    From left to right, the top row of figures displays the rank usage time series of the word 
    ``bling'' on Twitter,
    the price of the cryptocurrency Bitcoin, and a simple Gaussian random walk.  
    Below each time series we display parameter sweeps over combinations of $(\theta, W_{\max})$ for two kernel functions: one kernel given by the function of Eq.\ \ref{eq:cusp-kernel} and
another of the identical form but constructed by setting
$\kernel{S}(\tau|W,\theta)$ to the function given in Eq. 
\ref{eq:exp-kernel}.
The $\ell_1$ norms of the shock indicator function are nearly invariant across the values of the parameters $\theta$ for which we evaluated the kernels.
However, the shock indicator function does display dependence on the maximum window size $W_{\max}$, with large $W_{\max}$ associated with larger $\ell_1$ norm.
This is because a larger window size allows the DST to detect shock-like behavior over longer periods of time.
}
\label{fig:parameters_sweep}
\end{figure*}

\noindent
To aggregate deterministic behavior across all timescales of interest,
we define the shock indicator function as the function 
\begin{equation}\label{eq:sif}
	\mathrm{C}_{\kernel{S}}(t|\theta) = 
	\sum_W \mathrm{C}_{\kernel{S}}(t, W|\theta)p(W|\theta),
\end{equation}
for all windows $W$ considered.
The function $p(W|\theta)$ is a probability mass function that encodes prior beliefs about the importance of particular values of $W$.
For example, if we are interested primarily 
in time series that display shock- or shock-like behavior that usually lasts for approximately one month, we might 
specify $p(W|\theta)$ to be sharply peaked about $W = 28$ days.
Throughout this work we take an agnostic view on all possible window widths and 
so set $p(W|\theta) \propto 1$, reducing our analysis to a strictly maximum-likelihood based approach.
Summing over all values of the shocklet parameter $\theta$ defines the shock indicator function,
\begin{align}
		\mathrm{C}_{\kernel{S}}(t) &= 
		\sum_{\theta} \mathrm{C}_{\kernel{S}}(t|\theta) p(\theta) \\
		&= \sum_{\theta, W} \mathrm{C}_{\kernel{S}}(t, W |\theta)p(W|\theta) p(\theta).
\end{align}
In analogy with $p(W\theta)$, the function $p(\theta)$ is a probability density 
function describing our prior beliefs about the importance of various values of $\theta$.
As we will show later in this section, and graphically in Fig.\ \ref{fig:parameters_sweep}, the shock indicator function is relatively insensitive to choices of $\theta$ possessing a nearly-identical $\ell_1$ norm for wide ranges of $\theta$ and different functional forms of $\kernel{S}$.
\medskip

\noindent
After calculation, we normalize $\mathrm{C}_{\kernel{S}}(t)$ so that it again integrates to zero and has 
$\max_t \mathrm{C}_{\kernel{S}}(t) - \min_t \mathrm{C}_{\kernel{S}}(t) = 2$.
The shock indicator function is used to find windows in which the time series displays anomalous shock- or shock-like behavior. These windows are defined as 
\begin{equation}\label{eq:window-definition}
	\left\{t\in [0, T]:\text{intervals where } \mathrm{C}_{\kernel{S}}(t) \geq s \right\}.
\end{equation}
where the parameter $s > 0$ sets the sensitivity of the detection.
\medskip

\noindent
The DST is relatively insensitive to quantitative changes to its functional parameterization; 
it is a qualitative tool to highlight time periods of unusual events in a time series.
In other words, it does not detect statistical anomalies 
but rather time periods during which the time series appears
to take on certain qualitative characteristics without being too sensitive to a particular functional form.
We analyzed two example sociotechnical time series---the rank of the word ``bling'' on Twitter (for reasons we will discuss presently)--- and the price time series of Bitcoin (symbol BTC)
\cite{nakamoto2008bitcoin}, the most actively-used cryptocurrency
\cite{al2014investigating}, and of one null model, a pure random walk.
For each time series, we computed the shock indicator function using two kernels, each of which had a different functional form (one kernel given by the function of Eq.\ \ref{eq:cusp-kernel} and
one of the identical form but constructed by setting
$\kernel{S}(\tau|W,\theta)$ to the function given in Eq. 
\ref{eq:exp-kernel}), and evaluating each kernel over a wide range of its parameter 
$\theta$. We also vary the maximum window size from $W = 100$ to $W = 1000$ to explore the sensitivity of the shock indicator function to this parameter.
We display the results of this comparative analysis in Fig.\ \ref{fig:parameters_sweep}.
For each time series, we plot the $\ell_1$ norm of the shock indicator function for each $(\theta, W)$ combination.
We find that, as stated earlier in this section, the shock indicator function is relatively insensitive to both functional parameterization and value of the parameter $\theta$; for any fixed $W$, the $\ell_1$ norm of the shock indicator function barely changed regardless of the value of $\theta$ or choice of $\kernel{\cdot}$.
However, the maximum window size does have a notable effect on the magnitude of the shock indicator function; higher values of $W$ are associated with larger magnitudes.
This is a reasonable finding, since higher maximum $W$ means that the DST is able to capture shock-like behavior that occurs over longer timespans and hence may have values of higher magnitude over longer periods than for comparatively lower maximum $W$.
\medskip

\noindent
That the shock indicator function is a relative quantity is both beneficial and problematic. The utility of this feature
is that the dynamic behavior of time series derived from systems of widely-varying time and length scales can be 
directly compared;
while the rank of a word on Twitter and---for example---the volume of trades in an equity security are entirely different phenomena 
measured in 
different units, their shock indicator functions are unitless and share similar properties. 
On the other hand, the Shock Indicator Function carries with it no notion of dynamic range.
Two time series $x_t$ and $y_t$ could have identical shock indicator functions but have spans differing by many orders of magnitude, i.e.,
$\text{diam } x_t \equiv \max_t x_t - 
\min_t x_t \gg \text{diam } y_t$.
(In other words, the diameter of a time series in interval $I$ is just the dynamic range of the time series over that interval.)
We can directly compare time series inclusive of their dynamic range by computing a weighted version of the shock indicator function, $\mathrm{C}_{\mathcal{K}}(t) \Delta x(t)$, which we term the weighted shock indicator function (WSIF). 
A simple choice of weight is
\begin{equation}\label{eq:window-weight}
	\Delta x(t) = \diam\displaylimits_{t \in [t_b, t_e]}x_t,
\end{equation}
where $t_b$ and $t_e$ are the beginning and end times of a particular window.
We use this definition for the remainder of our paper, but one could easily imagine using other weighting functions, e.g., maximum percent change (perhaps applicable for time series hypothesized to increment geometrically instead of arithmetically).
\medskip

\begin{figure*}
\centering
\includegraphics[width=\textwidth]{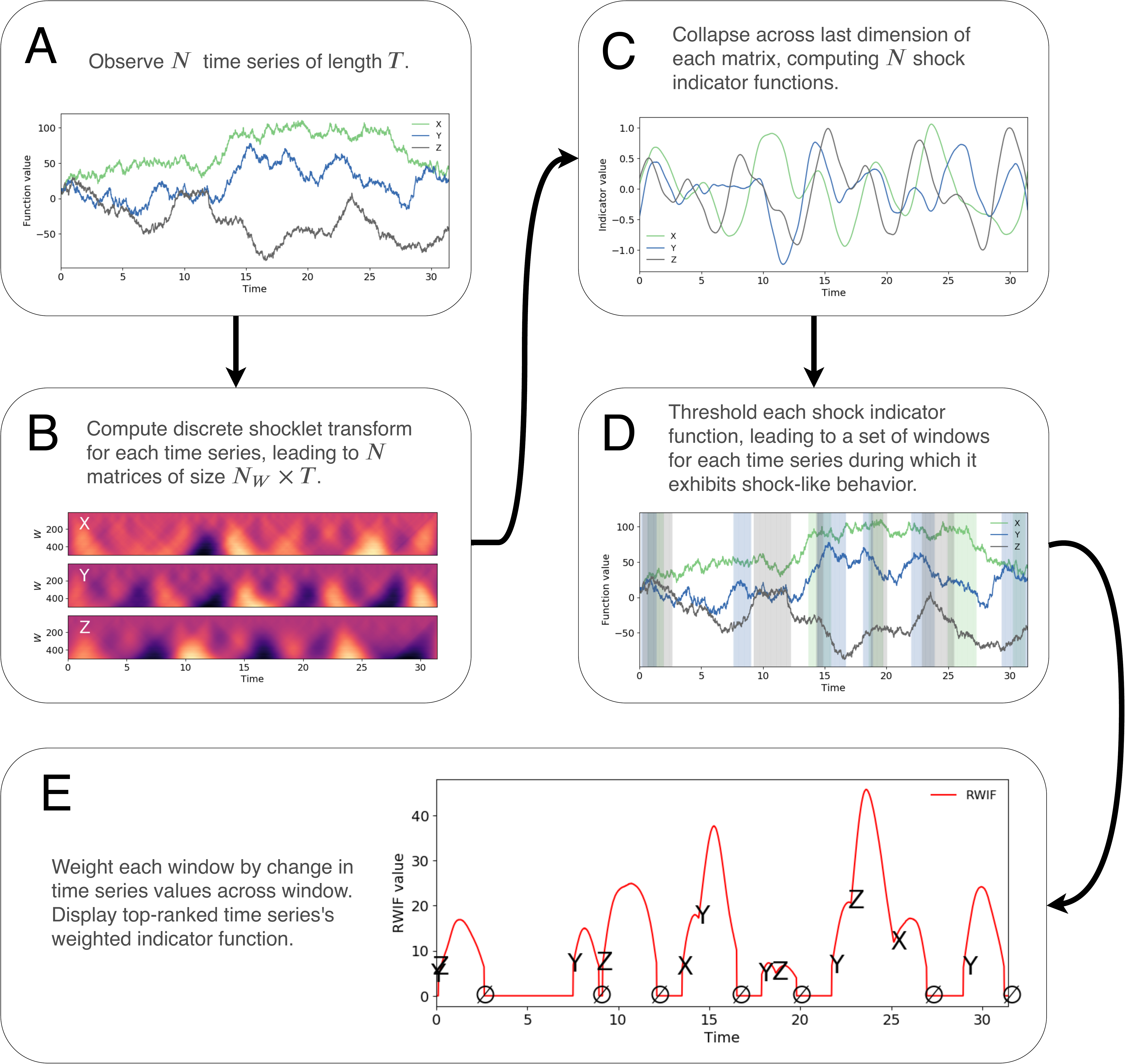}
\caption{
The Shocklet Transform And Ranking (STAR) algorithm combines the discrete shocklet transform (DST) with a series of transformations that yield intermediate results, 
such as the cusp indicator function (panel (C) in the figure) and windows during which each univariate time series displays shock-like behavior (panel (D) in the figure). Each of these intermediate results is useful in its own right, as we show in Sec.\ \ref{sec:results}. We display the final output of the STAR algorithm, a univariate indicator that condenses information about which of the time series 
exhibits the strongest shock-like behavior at each point in time.
}
\label{fig:star-diagram}
\end{figure*}

\noindent
These final weighted shock indicator functions are the ultimate output of the shocklet transform and ranking (STAR) algorithm;
the weighting corresponds to the actual magnitude of the 
dynamics and constitutes the ``ranking'' portion of the algorithm,
while the weighting will only be substantially larger than zero if there existed intervals of time during which the time series exhibited shock-like behavior as indicated in Eq.\ \ref{eq:window-definition}.
We present a conceptual, bird's-eye view of the STAR algorithm (of which the DST is a core component) in Fig. \ref{fig:star-diagram}.
Though this diagram is lacking in technical detail, we have included it in an effort to provide a bird's-eye view of the entire STAR algorithm and to help orient the reader on the conceptual process underpinning the algorithm.
\medskip

\subsubsection{Algorithmic details: Comparison with existing methods}
\label{sec:comparison}

On a coarse scale, there are five nonexclusive categories of time series data mining tasks \cite{lin2007experiencing}:
similarity search (also termed indexing), clustering, classification, summarization, and anomaly detection. 
The STAR algorithm is a qualitative, shape-based, timescale-independent, similarity search algorithm. 
As we have shown in the previous section, the discrete shocklet transform (a core part of the overarching STAR algorithm) is \textit{qualitative}, meaning that it does not depend too strongly on values of functional parameters or even the functions used in the cross-correlation operation themselves, as long as the functions share the same qualitative dynamics (e.g., increasing rates of increase followed by decreasing rates of decrease for cusp-like dynamics); hence, it is primarily \textit{shape-based} rather than relying on the quantitative definition of a particular functional form.
STAR is \textit{timescale-independent} as it is able to detect shock-like dynamics over a wide range of timescales limited only by the maximum window size for which it is computed.
Finally, we believe that it is best to categorize STAR as a \textit{similarity search} algorithm as this seems to be the best-fitting label for STAR that is given in the five categories listed at the beginning of this section; STAR is designed for searching within sociotechnical time series for dynamics that are similar to the shock kernel in some way, albeit similar in a qualitative sense and over any arbitrary timescale, not functionally similar in numerical value and characteristic timescale.
However, it could also be considered a type of qualitative, shape-based anomaly detection algorithm because we are searching for behavior that is, in some sense, anomalous compared to a usual baseline behavior of many time series (though see discussion at the beginning of the anomaly detection subsection near the end of this section: STAR is an algorithm that can detect defined anomalous behavior, \textit{not} an algorithm to detect arbitrary statistical anomalies).
\medskip

\noindent
As such, we are unaware of any existing algorithm that satisfies these four criteria and believe that STAR represents an entirely new class of algorithms for sociotechnical time series analysis. 
Nonetheless, we now provide a detailed comparison of the DST with other algorithms that solve related problems, and in Sec.\ \ref{sec:tw-comparison} provide an in-depth quantitative comparison with another nonparametric algorithm (Twitter's anomaly detection algorithm) that one could attempt to use to extract shock-like dynamics from sociotechnical time series.
\medskip

\noindent
\textit{Similarity search} - here the objective is to find time series that minimize some similarity criterion between candidate time series and a given reference time series. Algorithms to solve this problem include 
nearest-neighbor methods (e.g., $k$-nearest neighbors \cite{yang2007efficient} 
or a locality-sensitive hashing-based method \cite{kale2014examination,driemel2017locality}),
the discrete Fourier and wavelet transforms 
\cite{keogh2000simple,wu2000comparison,popivanov2002similarity,chan2003haar}; and bit-, string-, and matrix-based
representations \cite{ratanamahatana2005novel,keogh2005hot,lin2007experiencing,yeh2016matrix}.
With suitable modification, these algorithms can also be used to solve time series clustering problems. 
Generic dimensionality-reduction techniques, such as singular value decomposition / principal components analysis 
\cite{eastman1993long,harris1997principal,lansangan2009principal}, can also be used for similarity search by searching through a dataset of lower dimension.
Each of these classes of algorithms differs substantially in scope from the discrete shocklet transform.
Chief among the differences is the focus on the entire time series.
While the discrete shocklet transform implicitly searches the time series for similarity with the kernel function at 
all (user-defined) relevant timescales and returns qualitatively-matching behavior at the corresponding timescale,
most of the algorithms considered above do no such thing; the user must break the time series
into sliding windows of length $\tau$ and execute the algorithm on each sliding window; 
if the user desires timescale-independence, they must then vary $\tau$ over a desired range.
An exception to this statement is Mueen's subsequence similarity search algorithm (MSS) \cite{mueenfastest},
which computes sliding dot products (cross-correlations)
between a long time series of length $T$ and a shorter kernel of length $M$ 
before defining a Euclidean distance objective for the similarity search task.
When this sliding dot product is computed using the fast Fourier transform, the computational complexity of this task is $\mathcal{O}(T \log T)$.
This computational step is also at the core of the discrete shocklet transform, but is performed for multiple
kernel function arrays (more precisely, for the kernel function resampled at multiple user-defined timescales).
Unlike the discrete shocklet transform, MSS does not subsequently compute an indicator function and does not 
have the self-normalizing property, while the matrix profile algorithm
\cite{yeh2016matrix} computes an indicator function of sorts (their ``matrix profile'') but is not
timescale-independent and is quantitative in nature; it does not search for a qualitative shape match as does the
discrete shocklet transform.
We are unaware of a similarity-search algorithm aside from STAR
that is both qualitative in nature and timescale-independent.
\medskip
 
\noindent
\textit{Clustering} - given a set of time series, the objective is to group them into groups, or clusters,
that are more homogeneous within each cluster than between clusters.
Viewing a collection of $N$ time series of length $T$ as a set of vectors in $\mathbb{R}^T$, 
any clustering method that can be effectively used on high-dimensional data has potential applicability to clustering time series.  Some of these general clustering methods include $k$-means and $k$-medians algorithms
\cite{seref2013mathematical,vlachos2003wavelet,goutte1999clustering}, 
hierarchical methods \cite{jiang2003dhc,rodrigues2006odac,rodrigues2008hierarchical},
and density-based methods \cite{jiang2003dhc,denton2005kernel,birant2007st,ccelik2011anomaly}.
There are also methods designed for clustering time series data specifically, such as
error-in-measurement models \cite{kumar2002clustering},
hidden Markov models \cite{oates1999clustering}, simulated annealing-based methods \cite{schreiber1997classification},
and methods designed for time series that are well-fit by particular classes of parametric models \cite{kalpakis2001distance,bagnall2004clustering,xiong2004time,frohwirth2008model}.
Although the discrete shocklet transform component of the STAR algorithm could be coerced into performing
a clustering task by using different kernel functions and elements of the reflection group, clustering is not the
intended purpose of the discrete shocklet transform or STAR more generally. 
In addition, none of the clustering methods mentioned 
replicate the results of the STAR algorithm. 
These clustering methods uncover groups of time series that exhibit 
similar behavior over their entire domain; application of clustering methods to time series subsequences carries leads to meaningless results \cite{keogh2005clustering}.
Clustering algorithms are also shape-independent in the sense that they cluster data into groups that share similar
features, but do not search for specific known features or shapes in the data.  
In contrast with this, when using the STAR algorithm we already have specified a specific shape---for example, the 
shock shape demonstrated above---and are searching the data across timescales for occurrences of that shape.
The STAR algorithm also does not require multiple time series in order to function effectively, differing from any 
clustering algorithm in this respect; a clustering algorithm applied to $N=1$ data points 
trivially returns a single cluster containing the single
data point.
The STAR algorithm operates identically on one or many time series as it treats each time series independently.
\medskip
 
\noindent
\textit{Classification} - classification is the canonical supervised statistical learning problem in which data $x_i$ is observed along with a discrete label $y_i$ that is taken to be a function of the data,
$y_i = f(x_i) + \varepsilon$; the goal is to recover an approximation to $f$ that precisely and accurately reproduces the labels for new data \cite{bagnall2017great}.
This is the category of time series data mining algorithms that least corresponds with the STAR algorithm. 
The STAR algorithm is unsupervised---it does not require training examples (``correct labels'') in order to find 
subsequences that qualitatively match the desired shape. 
As above, the STAR algorithm also does not require
multiple time series to function well, while (non-Bayesian) classification algorithms rely on multiple data points
in order to learn an approximation to $f$ \footnote{
	Bayesian classification algorithms can perform classification
based only on prior information, but this is also not similar to the STAR algorithm, since the STAR algorithm
is a maximum-likelihood method that by definition requires at least one time series to operate.
}.
\medskip

\noindent
\textit{Summarization} - since time series can be arbitrarily large and composed of many intricately-related features,
it may be desirable to have a summary of their behavior that encompasses the time series's
``most interesting'' features. 
These summaries can be numerical, graphical, or linguistic in nature. 
Underlying methodologies for time series summary tasks include wavelet-based approaches
\cite{gilbert2001surfing,ahmad2004summarizing}, 
genetic algorithms \cite{castillo2011multi,castillo2011linguistic},
fuzzy logic and systems \cite{kacprzyk2007linguistic,kacprzyk2008linguistic,kacprzyk2010approach},
and statistical methods \cite{li2009dynammo}.
Though intermediate steps of the STAR algorithm can certainly be seen as a time series summarization mechanism
(for example, the matrix computed by the DShT or the weighted shock indicator functions used in determinning
rank relevance of individual time series at different points in time), the STAR algorithm was not designed for 
time series summarization and should not be used for this task as it will be outperformed by essentially any 
other algorithm that was actually designed for summarization.
Any ``summary'' derived from the STAR algorithm will have utility only in summarizing segments of the time series the
behavior of which match the kernel shape, or in distinguishing segments of the time series that do have a 
similar shape as the kernel
from ones that do not. 
\medskip

\noindent
\textit{Anomaly detection} - if a ``usual'' model can be defined for the system under study, an anomaly detection
algorithm is a method that finds deviations from this usual behavior.
Before we briefly review time series anomaly detection algorithms and compare them with the STAR algorithm, 
we distinguish between two subtly different concepts: this data mining notion of anomaly detection, and the
physical or social scientific notion of anomalous behavior.
In the first sense, \textit{any} deviation from the ``ordinary'' model is termed an anomaly and marked as such. 
The ordinary model may not be a parametric model to which the data is compared; for example, it may be implicitly defined as the behavior that the data exhibits most of the time --- whether in the context of temporal or other, e.g. spatial or network, data \cite{chandola2009anomaly,gutierrez2019multi}.
In physical and social sciences, on the other hand, it may be observed that, given a particular set of laboratory or observational conditions, a material, state vector, or collection of agents exhibits phenomena that is 
anomalous when compared to a specific reference situation, even if this behavior is ``ordinary'' for the conditions under which the phenomena is observed.
Examples of such anomalous behavior in physics and economics include: spectral behavior of polychromatic waves that is very unusual compared to the spectrum of monochromatic waves (even though it is typical for polychromatic waves near points where the wave's phase is singular) \cite{gbur2001anomalous};
the entire concept of anomalous diffusion, in which diffusive processes with mean square displacement (autocovariance functions) scaling as $\langle r(t)\rangle
\sim t^{\alpha}$ are said to diffuse anomalously if 
$\alpha \not\approx 1$ (since $\alpha = 1$ is the scaling of the Wiener process's autocovariance function) \cite{plerou2000economic,jeon2011vivo}, even though anomalous diffusion is the rule rather than the exception in intra-cellular and climate dynamics, as well as financial market fluctuations;
and behavior that deviates substantially from the  ``rational expectations'' of non-cooperative game theory, even though such deviations are regularly observed among human game players \cite{palfrey1997anomalous, capra1999anomalous}. 
This distinction between algorithms designed for the task of anomaly detection and algorithms or statistical procedures that test for the existence of anomalous behavior, as defined here, is thus seen to be a subtle but significant difference.
The DST and STAR algorithm fall into the latter category: the purpose for which we designed the STAR algorithm is to extract windows of anomalous behavior as defined by comparison with a particular null qualitative time series model (absence of clear shock-like behavior), not to perform the task of anomaly detection writ large by indicating the presence of arbitrary samples or dynamics in a time series that does not in some way comport with the statistics of the entire time series.
\medskip

\noindent
With these caveats stated, it is not the case that there is no overlap between anomaly detection algorithms and algorithms that search for some physically-defined anomalous behavior in time series; in fact, as we show in Sec.\ \ref{sec:tw-comparison}, there is some significant convergence between windows of shock-like behavior indicated by STAR and windows of anomalous behavior indicated by Twitter's anomaly detection algorithm when the underlying time series exhibits relatively low variance.
Statistical anomaly detection algorithms typically propose a semi-parametric model or nonparametric test and confront data with the model or test; if certain datapoints are very unlikely under the model or exceed certain theoretical boundaries derived in constructing the test, then these datapoints are said to be anomalous. 
Examples of algorithms that operate in this way include:
Twitter's anomaly detection algorithm (ADV), which relies on generalized seasonal ESD test \cite{rosner1983percentage,vallis2014novel};
the EGADS algorithm, which relies on explicit time series models and outlier tests \cite{laptev2015generic};
time-series model and graph methodologies 
\cite{chan2005modeling,cheng2009detection};
and probabilistic methods \cite{qiu2012granger,akouemo2016probabilistic}.
Each of these methods is strictly focused on solving the first problem that 
we outlined at the beginning of this subsection:
that of finding points in one or more time series during which it exhibits behavior that deviates substantially from the ``usual'' or assumed behavior for time series of a certain class.
As we outlined, this goal differs substantially from the one for which we designed STAR: searching for segments of time series (that may vary widely in length) during which the time series exhibits behavior that is qualitatively similar to underlying deterministic dynamics (shock-like behavior) that we believe is anomalous when compared to non-sociotechnical time series. 

\section{Empirical results} \label{sec:results}

\subsection{Comparison with Twitter's anomaly detection algorithm}
\label{sec:tw-comparison}
Through the literature review in Sec.\ \ref{sec:theory} we have demonstrated that,
to our knowledge, there exists no algorithm that solves the same problem for which STAR was designed---to provide a qualitative, shape-based, timescale-independent measure of similarity between multivariate time series and a hypothesized shape generated by mechanistic dynamics.
However, there are existing algorithms designed for nonparametric anomaly detection that could be used to alert to the presence of shock-like behavior in sociotechnical time series, which is the application for which we originally designed STAR. 
One leading example of such an algorithm is Twitter's Anomaly Detection Vector (ADV) algorithm \footnote{
\href{https://github.com/twitter/AnomalyDetection}{https://github.com/twitter/AnomalyDetection}.
}.
This algorithm uses an underlying statistical test, seasonal-hybrid ESD, to 
test for the presence of outliers in periodic and nonstationary time series
\cite{rosner1983percentage,vallis2014novel}.
We perform a quantitative and qualitative comparison between the STAR and ADV 
to compare their effectiveness at the task for which we designed STAR---determining qualitative similarity between shock-like shapes over a wide range of timescales---and to contrast the signals picked up by each algorithm, which, as we show, differ substantially.
Before presenting results of this analysis, we note that this comparison is not entirely fair; though ADV is a state-of-the-art anomaly detection algorithm, it was not designed for the task for which we designed STAR, and so it is not exactly reasonable to assume that ADV would perform as well as STAR on this task. 
In an attempt to ameliorate this problem, we have chosen a quantitative benchmark for which our \textit{a priori} beliefs did not favor the efficacy of either algorithm.
\medskip

\noindent
As both STAR and ADV are unsupervised algorithms, we compare their quantitative performance by assessing their utility in generating features for use in a supervised learning problem.
Since the macro-economy is a canonical example of a sociotechnical system, we consider the problem of predicting the probability of a U.S.\ economic recession
using only a minimal set of indicators from financial market data.
Models for predicting economic recessions variously use only real economic indicators \cite{chauvet1998econometric,dueker2005dynamic,osterholm2012limited},
only financial market indicators \cite{hamilton1996stock,estrella1998predicting}, or a combination of real and financial economic indicators \cite{qi2001predicting,berge2015predicting}. 
We take an approach that is both simple and relatively granular, focusing on the ability of statistics of individual equity securities to jointly model U.S.\ economic recession probability. 
For each of the equities that was in the Dow Jones Industrial Average between 1999-07-01 to 2017-12-31 (a total of $K=32$ securities),
we computed both the DST (outputting the shock indicator function), 
STAR algorithm (outputting windows of shock-like behavior), 
and the ADV routine on that equity's volume traded time series (number of shares transacted),
which we sampled at a daily resolution for a total of $T = 6759$ observations for each security.
We then fit linear models of the form
\begin{equation}\label{eq:logit-ols}
   E\left[ \log \mathbf{\frac{p}{1-p}}\right] = \mathbf{X}\bm{\beta},
\end{equation}
where $p_t$ is the recession probability on day $t$ as given by the U.S.\ Federal Reserve (hence $\mathbf{p}$ is the length-$T$ vector of recession probabilities) \footnote{
Data is available at 
\href{https://fred.stlouisfed.org/series/RECPROUSM156N}
{\\ https://fred.stlouisfed.org/series/RECPROUSM156N}.
}.
When we the model represented by Eq.\ \ref{eq:logit-ols} using ADV or STAR as the algorithms generating features, 
the design matrix $\mathbf{X}$ is a binary matrix of shape $T \times (K + 1)$ with
entry $X_{tk}$ equal to one if the algorithm indicated an anomaly or shock-like behavior respectively in security $k$ at time $t$ and equal to zero if it did not (the $+1$ in the dimensionality of the matrix corresponds to the prepended column of ones that is necessary to fit an intercept in the regression). 
When we fit the model using the shock indicator function generated by the DST, the matrix $\mathbf{X}$ is instead given by the matrix with column $k$ equal to the shock indicator function of security $k$.
\begin{figure}[!h]
\centering
\includegraphics[width=\columnwidth]{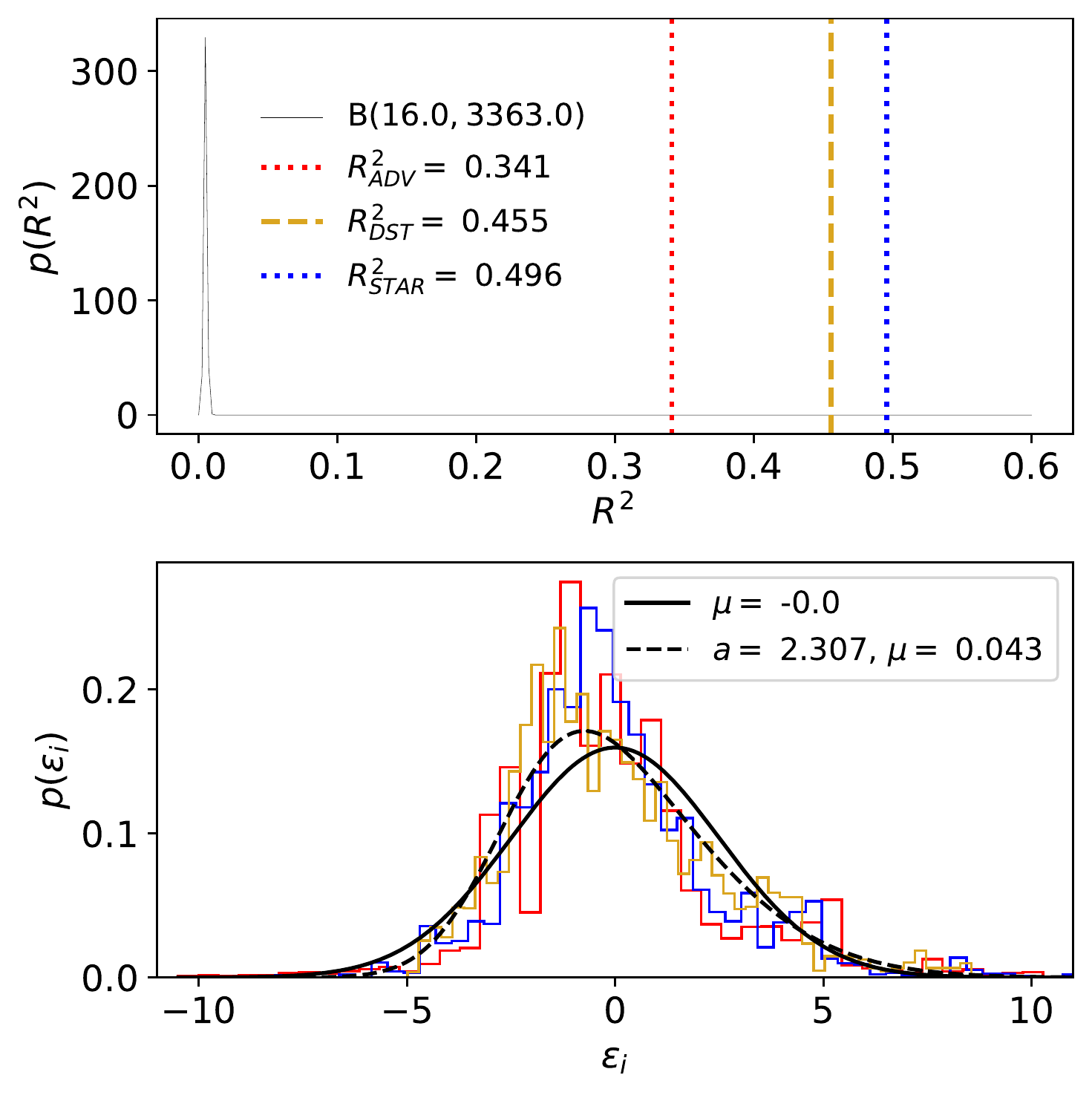}
\caption{ 
We modeled the log odds ratio of a U.S. economic recession using
three ordinary least squares regression models. 
Each model used one of the ADV method's anomaly indicator,
the shock indicator function resulting from the discrete shocklet transform, 
and the windows of shock-like behavior output by the STAR algorithm 
as elements of the design matrix.
The models that used features constructed by the DST or STAR outperformed the model that used features constructed by ADV as measured by both $R^2$ (displayed in the top panel) and model log-likelihood.
The black curve in the top panel displays the null distribution of $R^2$ under the assumption that no regressor (column of the design matrix) actually belongs to the true linear model of the data
\cite{cramer1987mean,carrodus1992exact}.
The lower panel displays the empirical probability distributions of the model residuals $\eps_i$. 
}
\label{fig:recession-r2-distributions}
\end{figure}
We evaluate the goodness of fit of these linear models using the proportion of variance explained ($R^2$) statistic; these results are summarized graphically in Fig.\ \ref{fig:recession-r2-distributions}.
The linear using ADV-indicated anomalies as features had $R^2_{ADV} = 0.341$,
while the model using the shock indicator function as columns of the design matrix had $R^2_{DST} = 0.455$ and 
the model using STAR-indicated shocks as features had $R^2_{STAR}= 0.496$.
This relative ranking of feature importance remained constant when we used model log-likelihood $\ell$ as the performance metric instead of $R^2$, with 
ADV, DST, and STAR respectively exhibiting $\ell_{ADV} = -16,278$,
$\ell_{DST} = -15,633$, and $\ell_{STAR} = -15,372$. 
Each linear model exhibited a distribution of residuals $\eps_t$ that did not drastically violate the zero-mean and distributional-shape assumptions of least-squares regression;
a maximum likelihood fit of a normal probability density to the empirical error probability distribution $p(
\eps_t)$ gave mean and variance as 
$\mu = 0$ to within numerical precision and 
$\sigma^2 \approx 6.248$,
while a maximum likelihood fit of a skew-normal probability density \cite{o1976bayes} to the empirical error probability distribution gave mean, variance, and skew as $\mu \approx 0.043$, 
$\sigma^2 \approx 6.025$, and $a \approx 2.307$.
Taken in the aggregate, these results constitute evidence to suggest that features generated by the DST and STAR algorithms are superior in the task of classifying time periods as belonging to recessions or not than are features derived from the ADV method.
\medskip

\begin{figure*}[!ht]
\includegraphics[width=\textwidth]{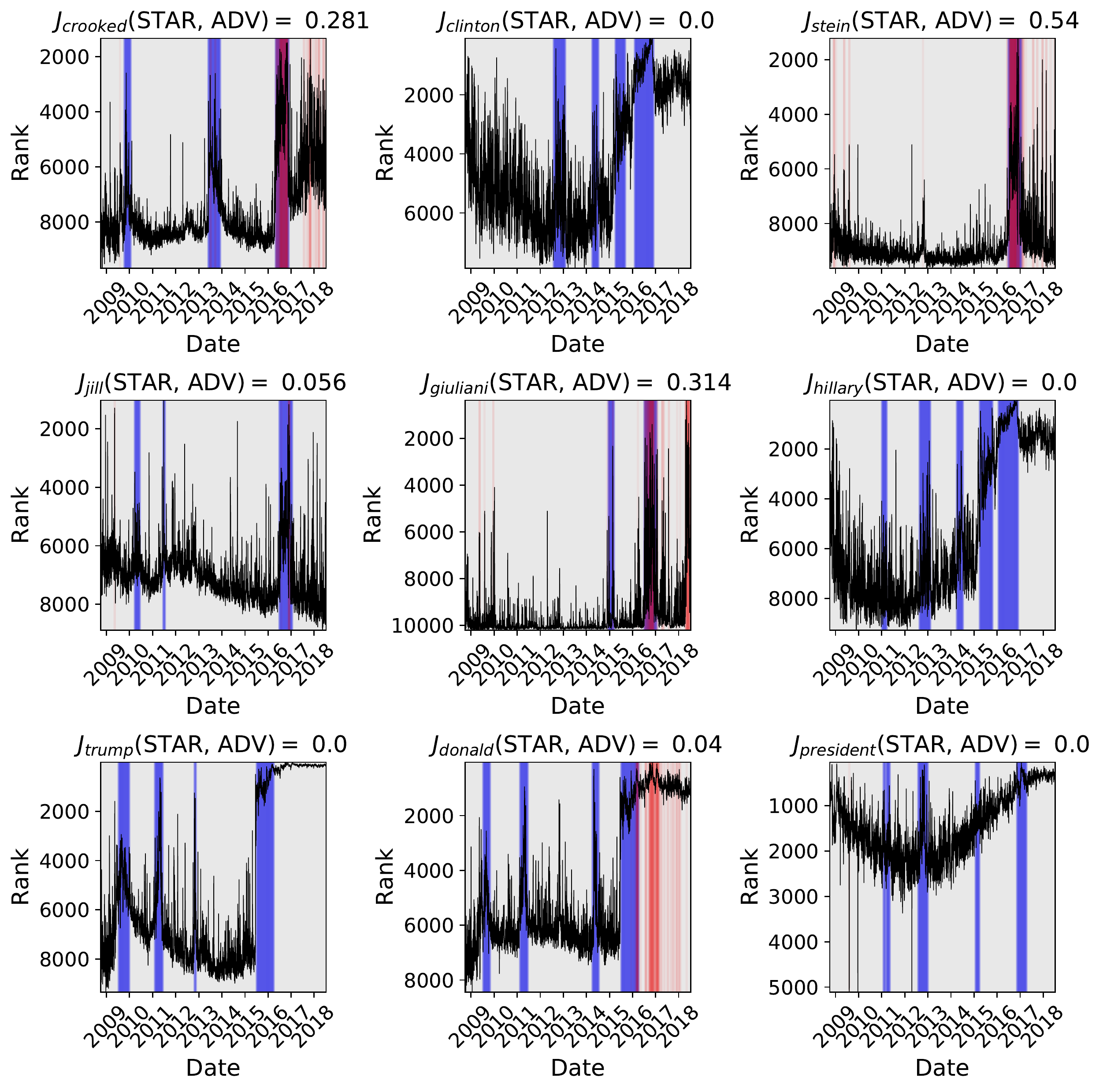}
	\caption{Comparison of STAR and Twitter's Anomaly Detection Vector (ADV) algorithm used for detecting phenomena in Twitter 1gram time series. 
	The Jaccard similarity coefficient is presented for each 1-gram and the region where events on detected are shaded for the respective algorithm. 
	Blue-shaded windows correspond with STAR windows of shock-like behavior, while red-shaded windows correspond with ADV windows of anomalous behavior (and hence purple windows correspond to overlap between the two).
	In general, ADV is most effective at detecting brief spikes or strong shock-like signals, whereas STAR is more sensitive to longer-term shocks and shocks that occur in the presence of surrounding noisy or nonstationary dynamic.
	ADV does not treat strong periodic fluctuations as anomalous by design; though this may or may not be a desirable feature of a similarity search or anomaly detection algorithm, it is certainly not a flaw in ADV but simply another differentiator between ADV and STAR. }
	\label{fig:crooked-clinton}
\end{figure*}

\noindent
As a further comparison of the STAR algorithm and ADV, we generated anomaly windows (in the case of ADV) and windows of shock-like behavior (in the case of STAR) for the usage rank time series of each of the 10,222 words in the LabMT dataset.
We computed the Jaccard similarity index for each word $w$ (also known as the intersection over union) between the set of STAR windows 
$\{I_i^{STAR}(w)\}_i$ and the set of ADV windows 
$\{I_i^{ADV}(w)\}_i$,
\begin{equation}
    J_w(STAR, ADV) =
    \frac{ \left( \bigcup_{i}I_i^{STAR}(w) \right) \cap \left( \bigcup_{i}I_i^{ADV}(w) \right)}
    {\bigcup_{j \in \{ STAR, ADV \}}\bigcup_i I_i^j(w) }.
\end{equation}
We display the word time series and ADV and STAR windows for a selection of words pertaining to the 2016 U.S.\ presidential election in Fig.\ 
\ref{fig:crooked-clinton}.
(These words display shock-like behavior in a time interval surrounding the election, as we demonstrate in the next section, hence our selection of them as examples here.)
A figure for each word that depicts the usage rank time series along with ADV and STAR-indicated windows is available at the authors' website 
\footnote{
\href{http://compstorylab.org/shocklets/all_word_plots/}{http://compstorylab.org/shocklets/all\_word\_plots/}
}.
\begin{figure}[!h]
\centering
\includegraphics[width=\columnwidth]{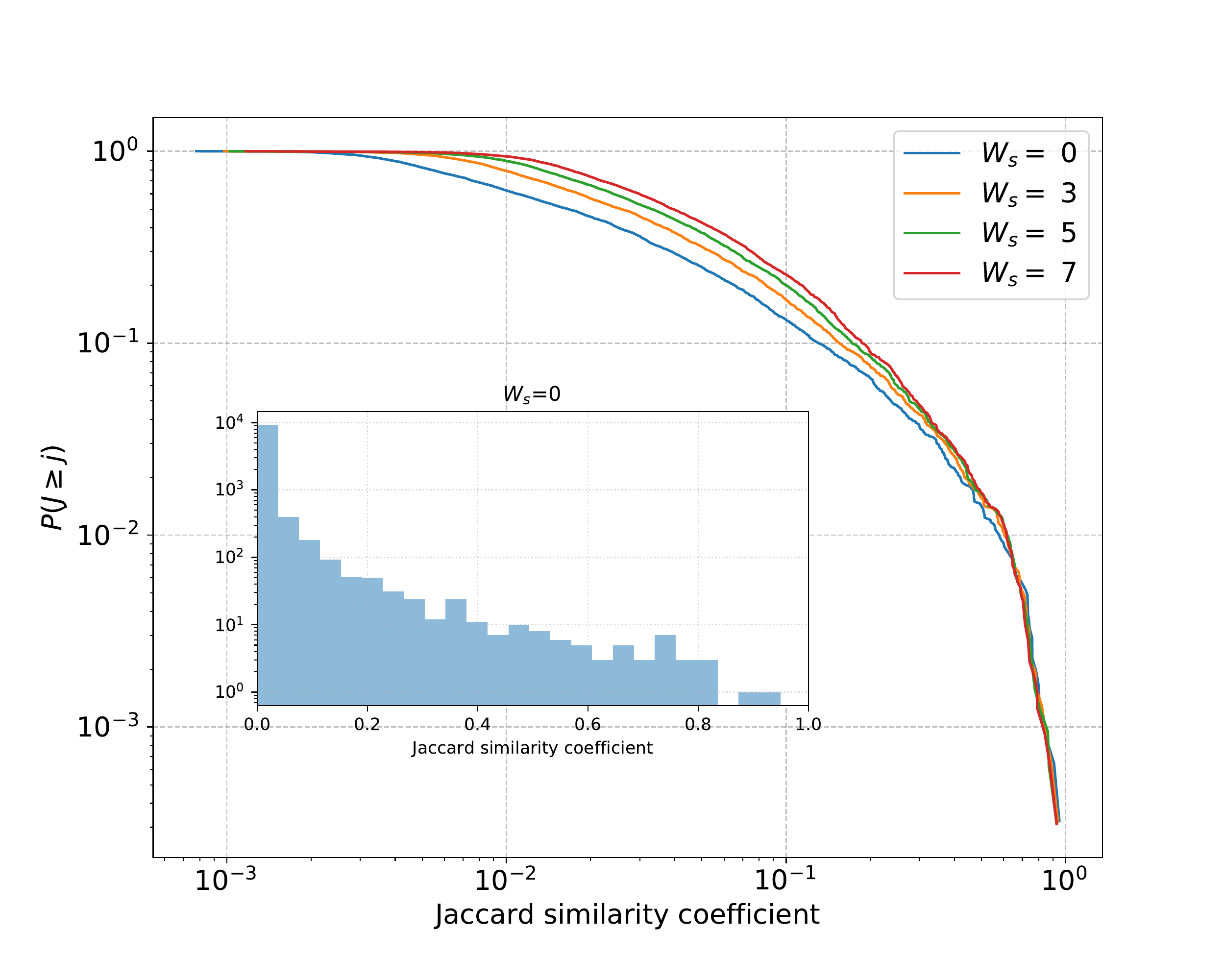}
\caption{ Complimentary cumulative distribution function (CCDF) of Jaccard similarity coefficients for regions that Twitter's ADV and our STAR algorithm detect patterns or anomalies (see Fig. \ref{fig:crooked-clinton}). 
Window sizes are varied to include $W_s \in \{0,3,5,7\}$ (i.e. detections within $t_i\pm W_s$ are as part of the intersection). 
Time series with $J_{\textnormal{word}_i}=0$ are omitted from the CCDF.
The inset histogram shows the distribution of Jaccard similarity coefficients for $W_s=0$ (i.e. exact matches), $J=0$ time series are included. }
\label{fig:CCDF_STAR_ADV}
\end{figure}
We display the distribution of all Jaccard similarity coefficients 
in Fig.\ \ref{fig:CCDF_STAR_ADV}.
Most words have relatively little overlap between anomaly windows returned by ADV and windows of shock-like dynamics returned by STAR, but there are notable exceptions.
In particular, a review of the figures contained in the online index suggests that ADV's and STAR's windows overlap most when the shock-like dynamics are particularly strong and surrounded by a time series with relatively low variance; they agree the most when hypothesized underlying deterministic mechanics are strongest and the effects of noise are lowest. 
The pronounced spikes in the words ``crooked'' and ``stein'' in Fig.\ \ref{fig:crooked-clinton} are an example of this phenomenon.
However, when the time series has high variance or exhibits strong nonstationarity,
ADV often does not indicate that there are windows of anomalous behavior while STAR does indicate the presence of shock-like dynamics;
the panels of the words ``trump'', ``jill'', and ``hillary'' in Fig.\ \ref{fig:crooked-clinton} demonstrate these behaviors. 
\medskip

\noindent
Taken in the aggregate, these results suggest that a state-of-the-art anomaly detection algorithm, such as Twitter's ADV, and a qualitative, shape-based, timescale-independent similarity search algorithm, such as STAR, do have some overlapping properties but are largely mutually-complementary approaches to identifying and analyzing the behavior of sociotechnical time series. 
While ADV and STAR both identify strongly shock-like dynamics that occur when the surrounding time series has relatively low variance, their behavior diverges when the time series is strongly nonstationary or has high variance.
In this case, ADV is an excellent tool for indicating the presence of strong outliers in the data, while STAR continues to indicate the presence of shock-like dynamics in a manner that is less sensitive to the time series's stationarity or variance. 

\subsection{Social narrative extraction}
\begin{figure*}
\includegraphics[width=\textwidth]{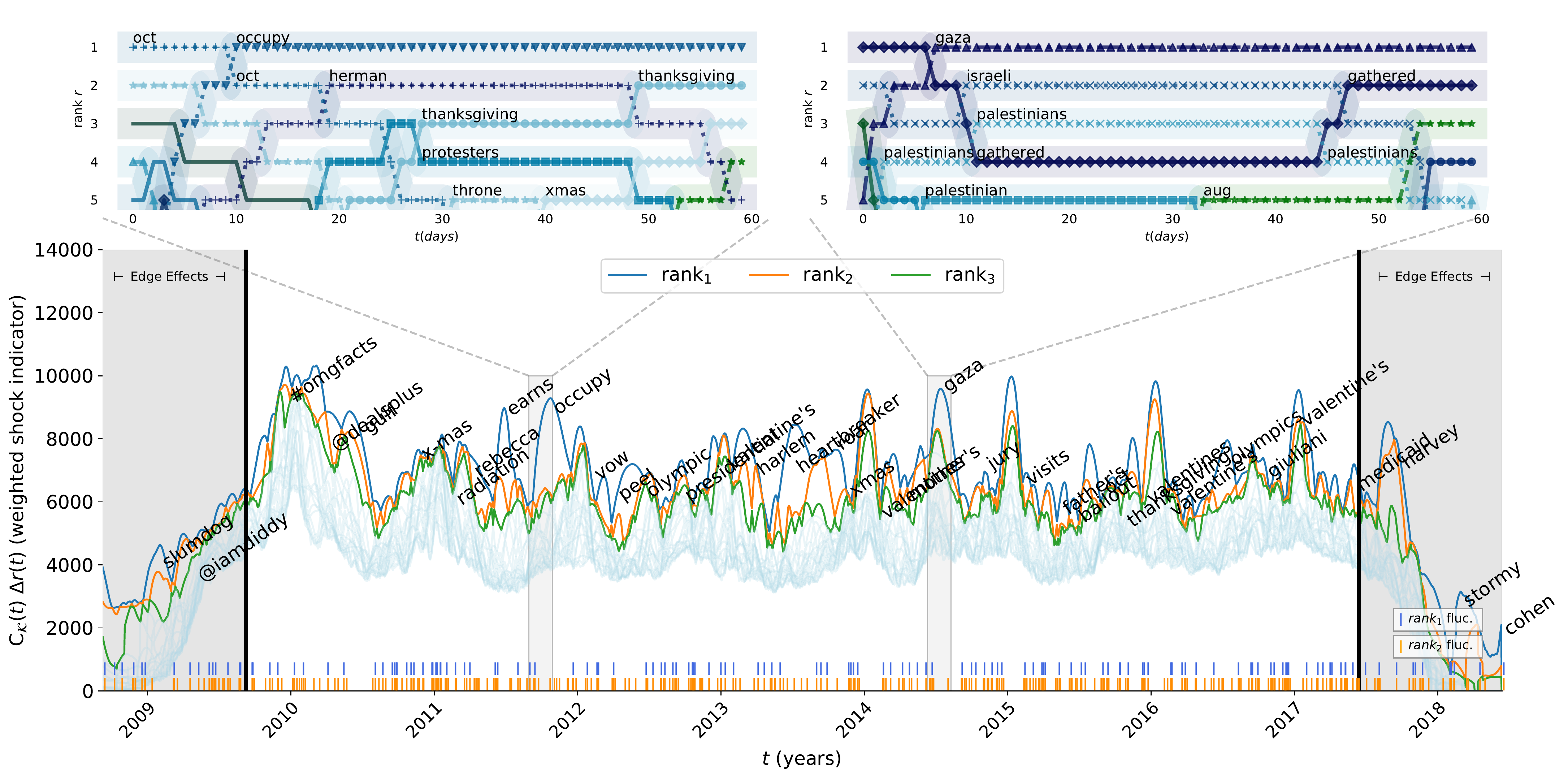}
	\caption{Time series of the ranked and weighted shock indicator function. 
	At each time step $t$, the weighted spike indicator functions (WSIF)
	are sorted so that the word with the highest WSIF corresponds to the top time series,
	the words with the second-highest WSIF corresponds to the second time series, and so on. Vertical ticks along the bottom mark fluctuations in the word occupying ranks 1 and 2 of WSIF values. Top panels present the ranks of WSIF values for words in the top 5 WSIF values in a given time step for the sub-sampled period of 60 days.
    An interactive version of this graphic is available 
	at the authors' webpage:  
	\www{http://compstorylab.org/shocklets/ranked_shock_weighted_interactive.html}.}
	\label{fig:topk-shock}
\end{figure*}

\begin{figure*}
\includegraphics[width=\textwidth]{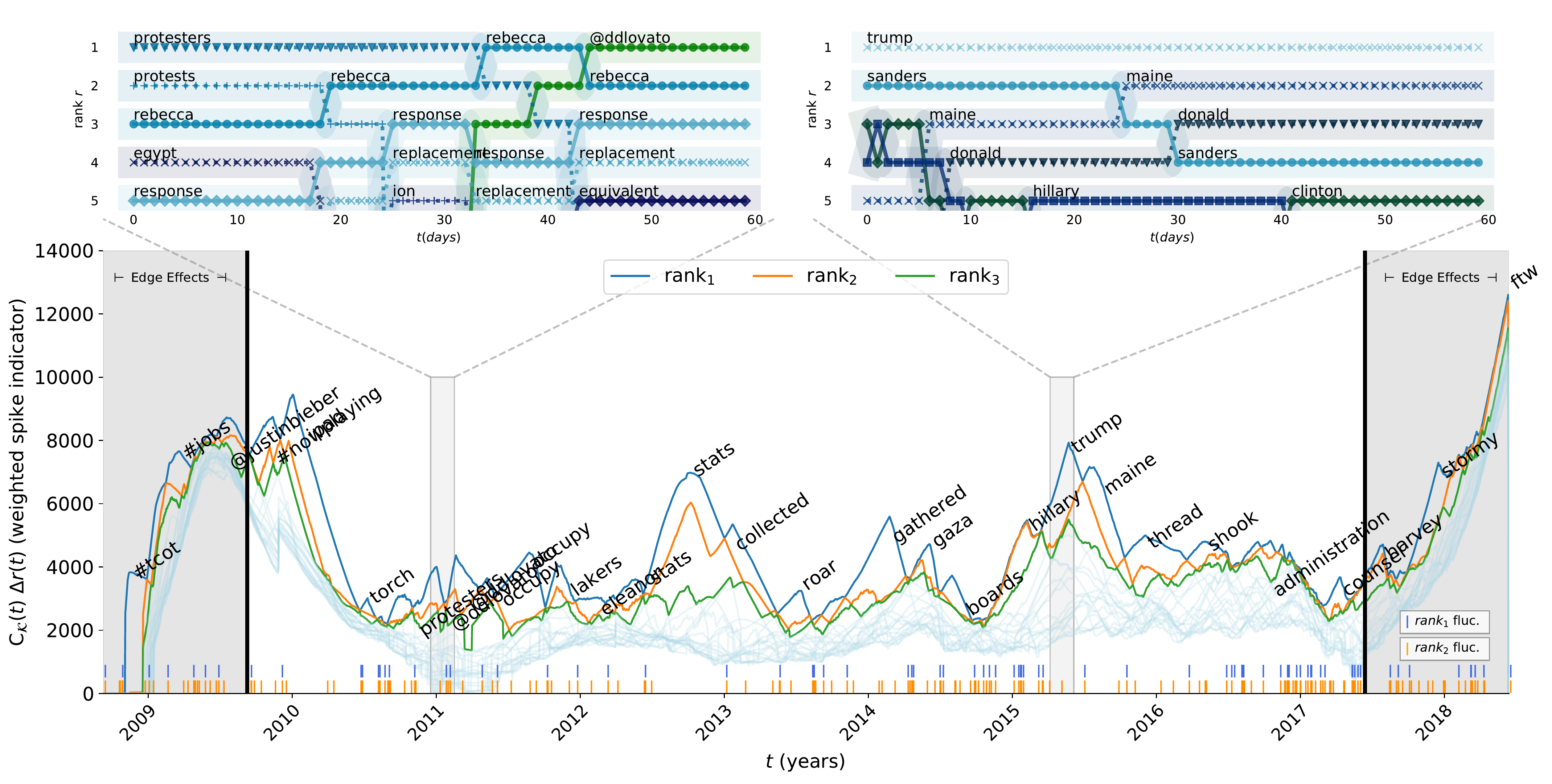}
	\caption{Time series of the ranked and weighted spike indicator function. 
	At each time step $t$, the weighted spike indicator functions (WSpIF)
	are sorted so that the word with the highest WSpIF corresponds to the top time series,
	the words with the second-highest WSpIF corresponds to the second time series, and so on. Vertical ticks along the bottom mark fluctuations in the word occupying ranks 1 and 2 of WSpIF values. Top panels present the ranks of WSpIF values for words in the top 5 WSpIF values in a given time step for the sub-sampled period of 60 days. The top left panel, demonstrates
    the competition for social attention between geopolitical
    concerns—street protests in Egypt--and popular artists
    and popular culture influence--Rebecca Black and Demi
    Lovato. The
    top  right  panel  displays  the  language  surrounding  the
    2016 U.S. presidential election immediately after Donald Trump
    announced his candidacy.
    An interactive version of this graphic is available 
	at the authors' webpage: 
	\www{http://compstorylab.org/shocklets/ranked_spike_weighted_interactive.html}.}
	\label{fig:topk-spike}
\end{figure*}

We seek both an understanding of the intertemporal semantic meaning imparted by windows of shock-like behavior indicated by the STAR algorithm
and a characterization of the dynamics of the shocks themselves.
We first compute the shock indicator and weighted shock indicator functions (WSIFs) for each of the 10,222 labMT words filtered from
the gardenhose dataset, described in section \ref{sec:data}, using a power kernel with $\theta=3$.
At each point in time, words are sorted by the value of their WSIF. 
The $j$-th highest valued WSIF at each temporal slice, when concatenated across time, 
defines a new time series. 
We perform this computation for the top ranked $k = 20$ words for the entire time under 
study. 
We also perform this process using the ``spike'' kernel of Eq.\ \ref{eq:spike-kernel}
and display each resulting time series in Fig.\ \ref{fig:topk-shock} (shock kernel) and Fig.\ \ref{fig:topk-spike} (spike kernel).
(We term the spike kernel as such because we have 
$\frac{\dee \kernel{Sp}(\tau)}{\dee \tau} = \delta(\tau)$ on the domain $[-W/2, W/2]$, the Dirac delta function;
its underlying mechanistic dynamics are completely static except for one point in time during which the system is driven by an ideal impulse function.)
The $j=1$ word time series is annotated with the corresponding word at relative maxima 
of order 40. (A relative maximum $x_s$ of order $k$ 
in a time series is a point that satisfies $x_s > x_t$ for all 
$t$ such that $|t - s| \leq k$.)
This annotation reveals a dynamic social narrative concerning popular events, social movements, and geopolitical 
fluctuation over the past near-decade.
Interactive versions of these visualizations are available on the authors' website \footnote{\href{http://compstorylab.org/shocklets/}{http://compstorylab.org/shocklets/}}. 
To further illuminate the often-turbulent dynamics of the top $j$ ranked weighted shock indicator functions, 
we focus on two particular 60-day windows of interest, denoted by 
shading in the main panels of Figs.\ \ref{fig:topk-shock} and \ref{fig:topk-spike}.
In Fig.\ \ref{fig:topk-shock},
we outline a period in late 2011 during which multiple events
 competed for collective attention:
 \begin{itemize}
 \item the 2012 U.S.\ presidential election (the word ``herman'', referring to Herman Cain, a presidential election contender);
 \item Occupy Wall Street protests (``occupy'' and ``protestors'');
 \item and the U.S.\ holiday of 
 Thanksgiving (``thanksgiving'')
 \end{itemize}
 Each of these competing narratives is reflected in the top-left 
 inset.
 In the top right inset, we focus on a time period during 
 which the most distinct anomalous dynamics corresponded to 
 the 2014 Gaza conflict with Israel 
 (``gaza'', ``israeli'', ``palestinian'', ``palestinians'', ``gathered'').
 In Fig.\ \ref{fig:topk-spike},
 we also outline two periods of time:
 one, in the top left panel,
demonstrates the competition for social attention 
between geopolitical concerns:
\begin{itemize}
    \item street protests in Egypt (``protests'', ``protesters'' ``egypt'', ``response'');
    \item and popular artists and popular culture
    (``rebecca'', referring to Rebecca Black, a musician, and ``@ddlovato'', referring to another musician, Demi Lovato).
\end{itemize}
In the top right panel we demonstrate that the most prominent 
dynamics during late 2015 are those of the language 
surrounding the 2016 U.S.\ presidential election immediately 
after Donald Trump announced his candidacy
(``trump'', ``sanders'', ``donald'', ``hillary'', ``clinton'', ``maine'').
\medskip

\noindent
We note that these social narratives uncovered by the STAR algorithm might not emerge if we used a different algorithm in an attempt to extract shock-like 
dynamics in sociotechnical time series. 
We have already shown (in the previous section) that at least one state-of-the-art anomaly detection algorithm is unlikely to detect abrupt, shock-like dynamics that occur in time series that are nonstationary or have high variance.
We display side-by-side comparisons of the indicator windows generated by each algorithm for every word in the LabMT dataset in the online appendix
(\href{http://compstorylab.org/shocklets/all_word_plots/}
{http://compstorylab.org/shocklets/all\_word\_plots/}).
A review of figures in the online appendix corresponding with words annotated in Figs.\ \ref{fig:topk-shock} and 
\ref{fig:topk-spike} 
provides evidence that an anomaly detection algorithm, such as ADV, may not necessarily capture the sane dynamics as does STAR.
We include selected panels of these figures in Appendix \ref{app:star-vs-adv}, displaying words corresponding with some peaks of the weighted shock and spike indicator functions.
(We hasten to note that this of course does not preclude the possibility that anomaly detection algorithms might indicate dynamics that are not captured by STAR.)

\subsection{Typology of local mechanistic dynamics}
\label{sec:typology-mech-dyn}
To further understand divergent dynamic behavior in word rank time series,
we analyze regions of these time series for which Eq.\ \ref{eq:window-definition} is satisfied---that is, where the value of the shock indicator function is greater than the sensitivity parameter.
We focus on shock-like dynamics since these dynamics qualitatively describe 
aggregate social focusing and subsequent de-focusing of attention mediated by the algorithmic substrate of the Twitter platform.
\begin{figure}
    \centering
    \includegraphics[width=\columnwidth]{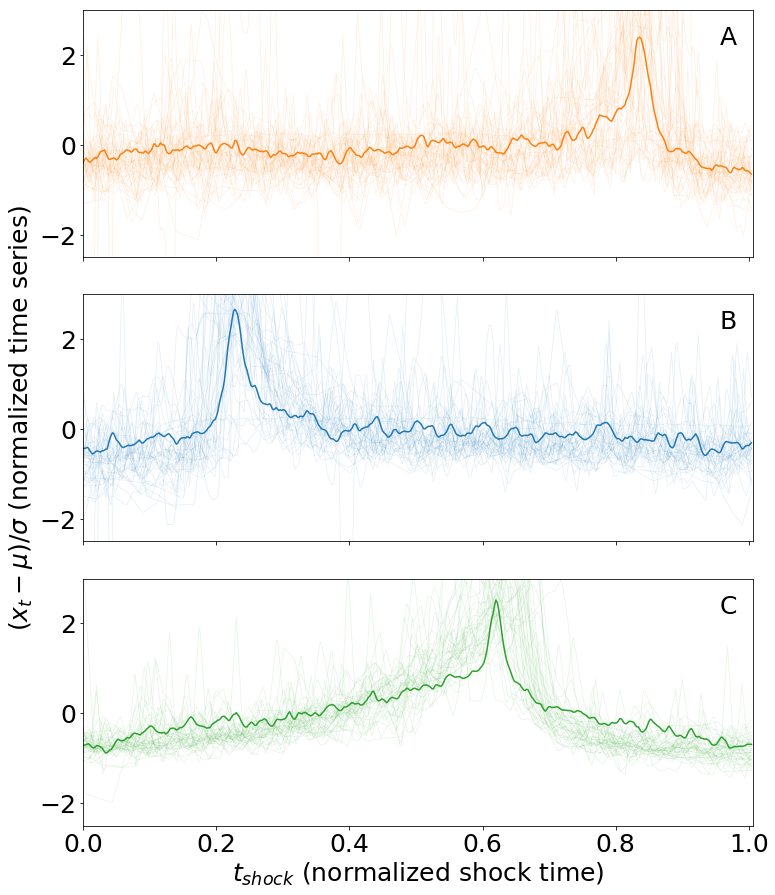}
    \caption{Extracted shock segments show diverse behavior corresponding to divergent social dynamics.
	We extract ``important'' shock segments (those that breach the top $k=20$ ranked weighted shock indicator at least once during the decade under study) and normalize them as described in Section \ref{sec:results}.
	We then find the densities of shock points $t^*_1$, 
	measured using the maxima of the within-window time series,
	and alternatively measured using the maxima of the (relative) shock indicator function. 
	We calculate relative maxima of these distributions and spatially-average shock segments whose maxima were closest to these
	relative maxima; we display these mean shock segments
	along with sample shock segments that are close to these mean shock segments in norm.
	We introduce a classification scheme for shock dynamics: 
	Type I (panel A) dynamics are those that display slow buildup and fast relaxation;
	Type II (panel B) dynamics, conversely, display fast (shock-like) buildup and slow 
	relaxation; and 
	Type III (panel C) dynamics are relatively symmetric. 
	Overall, we find that Type III dynamics are most common (40.9\%)
	among words that breach the top $k=20$ ranked weighted shock indicator function,
	while Type II are second-most common (36.4\%), followed by Type I (22.7\%).
	}
    \label{fig:cusp-shape-distribution}
\end{figure}  
We extract shock segments from the time series of all words 
	that made it into the top $j = 20$ ranked shock indicator functions at least once.
	Since shocks exist on a wide variety of dynamic ranges and timescales, 
	we normalize all extracted shock segments to lie on the time range $t_{\text{shock}} \in [0, 1]$
	and have (spatial) mean zero and variance unity.
	Shocks have a focal point about their maxima by definition, but in the context of stochastic 
	time series (as considered here), the observed maximum of the time series may not be the ``true'' maximum 
	of the hypothesized underlying deterministic dynamics. 
	Shock points---hypothesized deterministic maxima---of the extracted shock segments were thus determined by 
	two methods: The maxima of the within-window time series, 
	\begin{equation}
		t^*_1 = \argmax\displaylimits_{t_{\text{shock}} \in [0,1]} x_{t_{\text{shock}}};
	\end{equation}
	and the maxima of the time series's shock indicator function,
	\begin{equation}
		t^*_2 = \argmax\displaylimits_{t_{\text{shock}} \in [0,1]} \mathrm{C}_{\kernel{S}}(t_{\text{shock}}). 
	\end{equation}
We then computed empirical probability density functions of $t_1^*$ and $t_2^*$ across all words in the LabMT dataset.
While the empirical distribution of $t^*_1$ is uni-modal, the corresponding empirical distribution of $t^*_2$ demonstrated
clear bi-modality with peaks in the first and last quartiles of normalized time.
To better characterize these maximum \textit{a posteriori} (MAP)
estimates, we sample those shock segments $x_t$ the maxima of which are 
temporally-close to the MAPs and calculate spatial means of these samples, 
\begin{equation}
	\langle x_{t_{\text{shock}}} \rangle_n = \frac{1}{|\mathcal{M}|} \sum_{n \in \mathcal{M}} x_{t_{\text{shock}}}^{(n)},
\end{equation}
where 
\begin{equation}\label{eq:close-in-time}.
	\mathcal{M} = \left\{n: 
	\bigg |\argmax\displaylimits_{t_{\text{shock}} \in [0, 1]}  x_{t_{\text{shock}}}^{(n)}- t^*\bigg| < \eps \right \}.
\end{equation}
The number $\eps$ is a small value which we set here to $\eps = 10 / 503$ \footnote{
	This value comes from an arbitrary but small number of indices (five) we allow a shock segment to vary
	($\pm$)
	about the index of the MAP estimate of the distributions of shock points, each of which can be considered as 
	multinomial distributions supported on a 503-dimensional 
	vector space. 
	The number 503 is the dimension of each shock segment after time normalization since 
	the longest original shock segment in the labMT dataset was 503 days.
}.
We plot these curves in Fig.\ \ref{fig:cusp-shape-distribution}.
Shock segments that are close in spatial norm to the $\langle x_{t_{\text{shock}}} \rangle_n$---that is, 
shock segments $x_{t_{\text{shock}}}$ that satisfy
\begin{equation}\label{eq:close-in-space}
	\left\lVert x_{t_{\text{shock}}} - \langle x_{t_{\text{shock}}} \rangle_n \right\rVert_{1} 
	\leq 
	F_{\left\lVert x_s - \langle x_{t_{\text{shock}}} \rangle_n \right\rVert_{1} }^{\leftarrow}(0.01),
\end{equation}
where $F^{\leftarrow}_{Z}(q)$ is the quantile function of the random variable $Z$---are plotted in thinner curves.  
From this process, three distinct classes of shock segments emerge, corresponding with the three relative maxima of the 
shock point distributions outlined above: 

\begin{itemize}
    \item[-] \textbf{Type I}: exhibiting a slow buildup (anticipation) followed by a fast relaxation;
    \item[-] \textbf{Type II}: with a correspondingly short buildup (shock) followed by a slow relaxation;
    \item[-] \textbf{Type III}: exhibiting a relatively symmetric shape.
\end{itemize} 

\begin{table*}
\centering
	\begin{tabular}{|c|p{5cm}|p{11cm}|} \hline
	\label{tab:cusp-dynamic-classes}
	\textbf{Classification} & \textbf{Shock shape} & \textbf{Words} \\ \hline
		Type I & Slow buildup, fast relaxation & rumble, veterans, dusty, labour, scattered, hampshire, \#tinychat, elected, ballot, selection, labor, entering, beam, phenomenon, voters, mamma, anonymity, republican, \#nowplaying, indictment, wages, conservatives, pulse, knee, grammy, essays, \#tcot, kentucky, fml, netherlands, jingle, valid, whitman, syracuse, dems, deposit, bail, tomb, walker, reader\\ \hline
		Type II & Fast buildup, slow relaxation & xbox, chained, yale, bombing, holocaust, 
		connecticut, \#tinychat, civilian, jill, turkish, tsunami, ferry, 
		\#letsbehonest, beam, agreement, riley, ethics, phenomenon, harriet, privacy,
		israeli, \#nowplaying, gun, dub, pulse, killings, herman, enormous, fbi, dmc, 
		searched, norman, joan, affected, arthur, sandra, radiation, army, walker, reader, \\ \hline
		Type III & Roughly symmetric & rumble, memorial, sleigh, veterans, costumes, greeks, britney, separated, father's, shark, 
		grammys, labor, costume, x-mas, bunny, commonwealth, clause, olympics, olympic, daylight, cyber, wrapping, rudolph, drowned, re-election \\ \hline
	\end{tabular}
	\caption{Words for which at least one shock segment was close in norm to a spatial mean shock segment as 
	detailed in Section \ref{sec:results}.
    We display the distributions of ``shock points''---hypothesized deterministic maxima of the noisy mechanistically-generated time series---in Fig. \ref{fig:cusp-shape-distribution}.
    Every word may also have several ``shock points'' where each point could corresponds to a different shock dynamics due to the way each word is used throughout its life span on the platform, hence a few of these examples (e.g.\ rumble, anonymity, \#nowplaying) appear in multiple categories.
    }
\end{table*} 

Words corresponding to these classes of shock segments differ in semantic context.
Type I dynamics are related to known and anticipated
societal and political events and subjects, 
such as:
\begin{itemize}
	\item``hampshire'' and ``republican'', concerning U.S.\ presidential 
primaries and general elections,
	\item ``labor'', ``labour'', and ``conservatives'', likely concerning U.K.\ general elections,
	\item
``voter'', ``elected'', and ``ballot'', concerning voting in general,
and 
	\item``grammy'', the music awards show.
\end{itemize}
To contrast, Type II (shock-like) dynamics describe events that are partially- or entirely-unexpected, 
often in the context of national or international crises, such as:
\begin{itemize}
	\item ``tsunami'' and ``radiation'', relating to the Fukushima Daichii tsunami and nuclear meltdown,
	\item ``bombing'', ``gun'', ``pulse'', ``killings'', and ``connecticut'', concerning acts of violence and mass
		shootings, in particular the Sandy Hook elementary school shooting in the United States;
	\item ``jill'' (Jill Stein, a 2016 U.S.\ presidential election competitor), ``ethics'', and ``fbi'', pertaining to surprising events surrounding the 2016 U.S.\ presidential
		election, and 
	\item ``turkish'', ``army'', ``israeli'', ``civilian'', and ``holocaust'', concerning international protests, 
		conflicts, and coups.
\end{itemize}
Type III dynamics are associated with anticipated events that typically re-occur and are discussed 
substantially after their passing, such as
\begin{itemize}
	\item ``sleigh'', ``x-mas'', ``wrapping'', ``rudolph'', ``memorial'', ``costumes'', ``costume'', 
		``veterans'', and ``bunny'', 
		having to do with major holidays, and 
	\item ``olympic'' and ``olympics'', relating to the Olympic games.
\end{itemize}
We give a full list of words satisfying the criteria given in Eqs.\ \ref{eq:close-in-time} and \ref{eq:close-in-space} in Table \ref{tab:cusp-dynamic-classes}.
We note that, though the above discussion defines and distinguishes three fundamental signatures of word rank shock 
segments, these classes are only the MAP estimates of the true distributions of shock segments, our empirical 
observations of which are displayed as histograms in Fig.\ \ref{fig:cusp-shape-distribution};
there is an effective continuum of dynamics that is richer, but more complicated, than our parsimonious description here.

\section{Discussion} \label{sec:discussion}

We have introduced a nonparametric pattern detection method, termed
the discrete shocklet transform (DST) for its particular application in 
extracting shock- and shock-like dynamics from noisy time series,
and demonstrated its particular 
suitability for analysis of sociotechnical data. 
Though extracted social dynamics display a continuum of behaviors, we have
shown that maximizing \textit{a posteriori} estimates of shock likelihood
results in three distinct classes of dynamics: anticipatory dynamics with long buildups and quick relaxations, 
such as political contests (Type I);
``surprising'' events with fast (shock-like) buildups and long relaxation times, 
examples of which are acts of violence, natural disasters, and 
mass shootings (Type II);
and quasi-symmetric dynamics, corresponding with 
anticipated and talked-about events such as holidays and major sporting events (Type III).
We analyzed the most ``important'' shock-like dynamics---those words that were one of the top-20 most significant at least once during the decade of study---and found that Type III dynamics were the most common among these words (40.9\%)
followed by Type II (36.4\%) and 
Type I (22.7\%).
We then showcased the discrete shocklet transform's effectiveness in extracting 
coherent intertemporal narratives from word usage data
on the social microblog Twitter,  
developing a graphical methodology for examining meaningful fluctuations
in word---and hence topic---popularity.
We used this methodology to create document-free nonparametric topic models, 
represented by pruned networks based on shock indicator similarity between two words and defining topics using the
networks' community structures. 
This construction, while retaining artifacts from its construction using
intrinsically-temporal data, presents topics possessing
qualitatively sensible semantic structure.

There are several areas in which future work could improve on and extend that
presented here. 
Though we have shown that the discrete shocklet transform is a useful tool 
in understanding non-stationary local behavior when applied to a variety of sociotechnical time series, there is reason to suspect that one can generalize
this method to essentially any kind of noisy time series 
in which it can be hypothesized that mechanistic local dynamics contribute  
a substantial component to the overall signal. 
In addition, the DST suffers from noncausality, as do all convolution or frequency-space transforms. In order to compute an accurate transformed signal at time $t$, information about time $t + \tau$ must be known to avoid edge effects or spectral effects such as ringing.
In practice this may not be an impediment to the DST's usage, 
since:
empirically the transform still finds ``important'' local 
dynamics, as shown in Fig.\ \ref{fig:topk-shock} 
near the very beginning (the words ``occupy'' and ``slumdog'' are 
annotated)
and the end (the words ``stormy'' and ``cohen'' are annotated)
of time studied.
Furthermore, when used with more frequently-sampled data the
lag needed to avoid edge effects may have decreasing length relative 
to the longer timescale over which users interact with the data.
However, to avoid the problem of edge effects entirely,
it may be possible to train a supervised learning algorithm to 
learn the output of the DST at time $t$ using only past (and possibly 
present) data.
The DST could also serve as a useful counterpart to phrase- and sentence-tracking algorithms such as MemeTracker \cite{kleinberg2003bursty,leskovec2009meme}.
Instead of applying the DST to time series of simple words, one could apply it to arbitrary $n$-grams (including whole sentences) or sentence structure pattern matches to uncover frequency of usage of verb tenses, passive/active voice construction, and other higher-order natural language constructs. 
Other work could apply the DST to more and different natural language data sources or other sociotechnical time series, such as asset prices, economic indicators, and election polls.

\vspace*{10px}  
\section*{Acknowledgements}
 
The authors acknowledge the computing resources provided by the Vermont Advanced Computing Core and financial support from the Massachusetts Mutual Life Insurance Company, and are grateful for web hosting assistance from Kelly Gothard and useful conversations with Jane Adams and Colin Van Oort.

\bibliography{cusplets}{}
\bibliographystyle{unsrt}
 
\newpage

\appendix

\section{Statistical details}\label{app:stats}

In this appendix we will outline some statistical details of the DST and STAR algorithm that are not necessary for a qualitative understanding of them, but could be useful for more in-depth understanding or efforts to generalize them.
\medskip

\begin{figure*}
	\includegraphics[width=\textwidth]{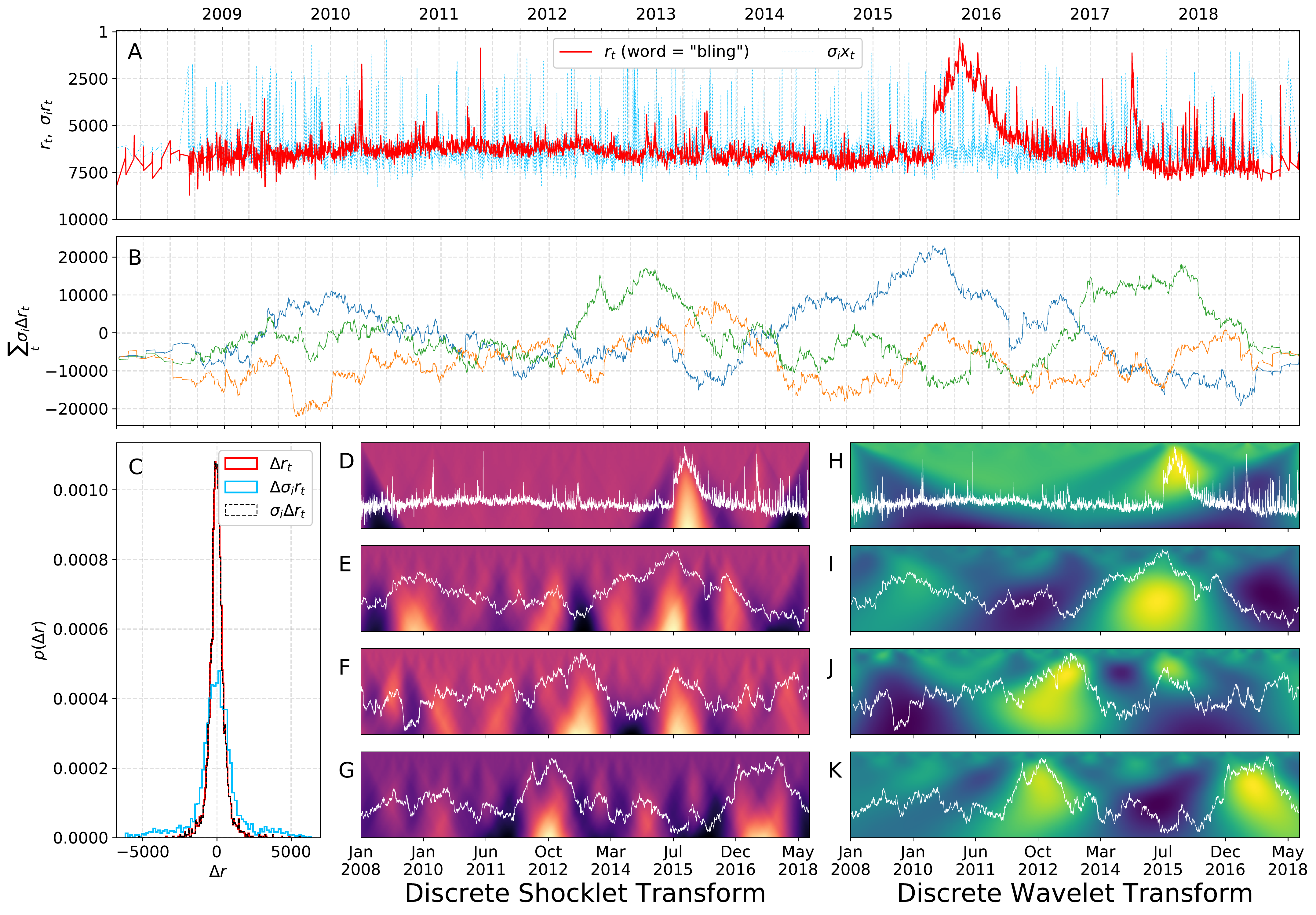}
	\caption{Intricate dynamics of sociotechnical time series.
	Sociotechnical time series can display intricate dynamics and extended periods of anomalous behavior.
	The red curve shows the time series of the ranks down from top of the word ``bling'' on Twitter.
	Until 2015/10/31, the time series presents as random fluctuation about a steady trend
	that is nearly indistinguishable from zero. 
	However, the series then displays a large fluctuation, increases rapidly, and then decays slowly after 
	a sharp peak.
	The underlying mechanism for these dynamics was the release of a popular song titled ``Hotline Bling'' by a musician known as ``Drake''. 
	Returns $\Delta r_t = r_{t + 1} - r_t$ are calculated and their histogram is 
	displayed in panel C.
	To demonstrate the qualitative difference of the ``bling'' time series from other time series 
	with an identical returns distribution, 
	elements of the symmetric group $\sigma_i \in \mathcal{S}_{T}$ are applied to the returns of the original series,
	$\Delta r_t \mapsto \Delta r_{\sigma_it}$, and the resultant noise is integrated and plotted as 
	$r_{\sigma_i t} = \sum_{t' \leq t}\Delta r_{\sigma_it}$.
	The bottom-left panel (C) displays time-decoupled probability distributions of the returns of the plotted time series.
	The distributions of $\Delta r_i$ and $\sigma \Delta r_i$ 
	are identical, as they should be, but the integrated series have
	entirely different 
	spectral behavior and dynamic ranges. 
	Panels [D-G] display the discrete shocklet transform of the original series 
	and the random walks $\sum_{t'\leq t} \Delta r_{\sigma_i t}$,
	showing the responsiveness of the DST to nonstationary 
	local dynamics and its insensitivity to dynamic range.
	The right-most column of panels [H-k] displays the discrete 
	wavelet transform of the original series
	demonstrating its comparatively less-sensitive nature to local
	anomalous dynamics.
	}\label{fig:symmetric-group-null-example_insane}
\end{figure*}

\noindent
We first give an illustrative example of how a sociotechnical time series can differ substantially from two null models of time series that have some similar statistical properties, displayed in Fig.\ \ref{fig:symmetric-group-null-example_insane} (a more information-rich version of Fig.\ \ref{fig:symmetric-group-null-example}, displayed in the main body), panels A and B. 
In panel A, we display an example sociotechnical time series in the red curve,  usage rank of the word ``bling'' within the LabMT subset of words on Twitter (denoted by $r_t$), 
and $\sigma r_t$, a randomly shuffled version of this time series. 
We denote $\sigma \in \mathcal{S}_T$, the symmetric group on $T$ elements, and draw $\sigma$ from the uniform distribution over $\mathcal{S}_T$.
It is immediately apparent that the structure of $r_t$ and $\sigma r_t$ are radically different in autocorrelation (both in levels and differences) and we do not investigate this admittedly-na\"{i}ve null model any further. 
\medskip

\noindent
We next consider a random walk null model constructed as follows: first differencing $r_t$ to obtain $\Delta r_t = r_t - r_{t-1}$, we 
apply random elements $\sigma_i \in \mathcal{S}_T$ and integrate, displaying
the resulting $r_{\sigma_i t}= \sum_{t' \leq t}\sigma_i\Delta r_t$ in panel C 
of Fig.\ \ref{fig:symmetric-group-null-example_insane}. 
Visual inspection (i.e., the ``eye test'') also demonstrates that these time series do not replicate the behavior displayed by the original $r_t$; they become negative, have a dynamic range that is almost an order of magnitude larger, and are more highly autocorrelated.
We contrast the results of the DST on $r_t$ and draws from this random walk null model in panels D -- G of Fig.\ \ref{fig:symmetric-group-null-example_insane}.
In panel D we display the DST of $r_t$, while in panels E -- G we display the DST of three random $\sigma_i r_t$.
The DSTs of the draws from the random walk model are more irregular that the DST of $r_t$, displaying many time-domain fluctuations between large positive values and large negative values.
In contrast, the DST of $r_t$ is relatively constant except near August of 2015, where it exhibits a large positive fluctuation across a wide range of $W$.
The underlying dynamics for this fluctuation were driven by the release of a popular song called ``Hotline Bling'' on July 31st, 2015.
\medskip

\noindent
As a couterpoint to the DST, we computed the discrete wavelet transform (DWT) of $r_t$ and the same $\sigma_i r_t$.
We computed the wavelet transform using the Ricker wavelet, 
\begin{equation}
    \psi(\tau,W) = \frac{2}{ \sqrt{3W\pi^{1/2}} }
    \left[1 - \left(\frac{\tau}{W}\right) \right]e^{-\tau^2/(2W^2)}.
\end{equation}
We chose to compare the DST with the DWT because these transforms are very similar in many respects: 
they both depend on two parameters (a location parameter $\tau$ and a scale parameter $W$); 
they both output a matrix of shape $T \times N_W$ ($N_W$ rows, 
one for each value $W$, and $T$ columns, one for each value of $\tau$). 
There are some key difference between these transforms, however. 
The ``kernels'' of the wavelet transform---the kernels---have unique properties not shared by our shock-like kernels: wavelets $\psi(t)$ are defined on all of $\mathbb{R}$,  
satisfy $\lim_{t \rightarrow \pm \infty} \psi(t) = 0$, and are orthonormal.
Our shock-like kernels do not satisfy any of these properties; they are defined on a finite interval $[-W/2, W/2]$, do not vanish at the endpoints of this interval, and are not orthogonal functions. Hence, differences in the DST and DWT of a time series are due primarily to the choice of convolution function---shock-like kernel in the case of the DST and wavelet in the case of the DWT. 
We display the DWT of $r_t$ and the same $\sigma_i r_t$ in panels H -- K of Fig.\ \ref{fig:symmetric-group-null-example_insane}.
Comparing these transforms with the DSTs displayed in panels D -- G, we see that the DST has increased time-localization over the DWT in time intervals during which the time series exhibit shock-like dynamics. 
As we note in Sec.\ \ref{sec:tw-comparison} (there when comparing STAR to Twitter's ADV anomaly detection algorithm), this observation should not be construed as equivalent to the statement that the DST is in some way superior to the DWT or should supersede the DWT for general time series processing tasks; rather, it is evidence that the DST is a superior transform than the DWT for the purpose of finding shock-like dynamics in sociotechnical time series---a task for which it was designed and the DWT was not. 
\medskip

\noindent
We finally note an analytical property of the DST that, while likely not useful in practice, is a fact that should be recorded and may be useful in constructing theoretical extensions of the DST. 
The DST is defined in Eq.\ \ref{eq:discrete-shocklet-transform}, which we record here for ease in reference:
\begin{equation}
    \mathrm{C}_{\kernel{\cdot}}(t, W| \theta)
    =\sum_{-\infty}^\infty x(t + \tau)\kernel{\cdot}(\tau|W,\theta),
\end{equation}
defined for each $t$.
The function $\kernel{\cdot}$ is the shock kernel that is non-zero on $\tau \in [-W/2 + t, W/2 + t]$. 
For $t \in [-T, T]$, this can be rewritten equivalently as 
\begin{equation}
     \mathbf{C}_{\kernel{\cdot}}(W |\theta)
      = \bm{K}(W |\theta)\bm{x},
\end{equation}
where $\bm{K}(W | \theta)$ is a $(2T + 1) \times (2T + 1)$ $W$-diagonal matrix,
$\mathbf{C}_{\kernel{\cdot}}(W |\theta)$ is the $W$-th row of the cusplet transform matrix, and $\bm{x}$ is the entire time series $x(t)$ considered as a vector in $\mathbb{R}^{2T + 1}$.
The matrix $\bm{K}(W | \theta)$ is just the convolution matrix corresponding to the cross-correlation operation with $\kernel{\cdot}$. 
If $\bm{K}(W|\theta)$ is invertible, then it is clear that
\begin{equation}
   \bm{x} =  \bm{K}(W |\theta)^{-1}\mathbf{C}_{\kernel{\cdot}}(W |\theta),
\end{equation}
for any $1 < W < T$ and hence also 
\begin{equation}
    \bm{x} = \frac{1}{N_{W}}\sum_W \bm{K}(W |\theta)^{-1}\mathbf{C}_{\kernel{\cdot}}(W |\theta).
\end{equation}
This is an inversion formula similar to the inversion formulae of overcomplete transforms such as the DWT and discrete chirplet transform.

When $T \rightarrow \infty$ (that is, when the signal $x(t)$ is turned on in the infinite past and continues into the infinite future), this equation becomes the formal operator equation
\begin{equation}
    \mathrm{C}_{\kernel{\cdot}}(t, W|\theta)
      = \bm{K}(W |\theta)[x(t)], 
\end{equation}
and hence (as long as the operator inverses are well-defined),
\begin{equation}
    x(t) = \frac{1}{N_W}\sum_W \bm{K}(W |\theta)^{-1}[\mathrm{C}_{\kernel{\cdot}}(t, W|\theta)].
\end{equation}
These inversion formulae are, in our estimation, of relatively little utility in practical application. 
Whereas inverting a wavelet transform is a common task---it may be desirable to decompress an image that is initially compressed using the JPEG 2000 algorithm, which uses the wavelet transform for compact representation of the image---we estimate the probability of being presented with some arbitrary shocklet transform and needing to recover the original signal from it to be quite low; the shocklet transform is designed to amplify features of signals to which we already have access, not to recreate time-domain signals from their representations in other domains.

\begin{figure*}[!th]
\centering
	\includegraphics[width=\textwidth]{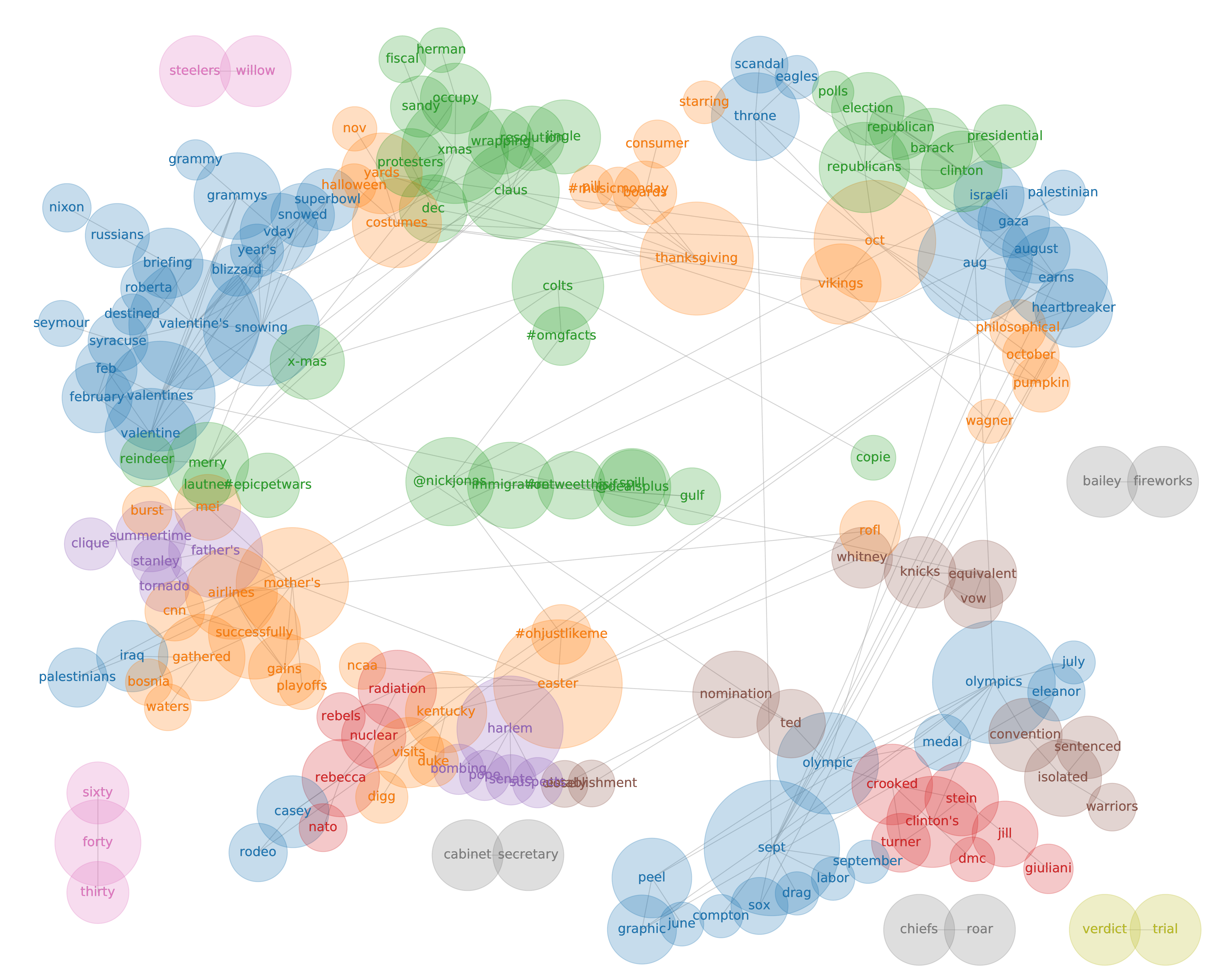}
	\caption{Topic network inferred from weighted shock indicator functions.
	At each point in time, words are ranked according to the value of their weighted shock
	indicator function and the top $k$ words are taken and linked pairwise for an 
	upper bound of $\binom{k}{2}$ additional edges in the network; if the edge between words $i$ and $j$ already exists, 
	the weight of the edge is incremented.
	The edge weight increment at time $t$ is given by $w_{ij,t} = \frac{R_{i,t} + R_{j,t}}{2}$,
	the average of the weighted shock indicator for words $i$ and $j$, 
	with the total edge weight thus given by $w_{ij} = \sum_t w_{ij,t}$.
	After initial construction, the backbone of the network is extracted using the method of Serrano \textit{et al.}
	\cite{serrano2009extracting}.
	The network is pruned further by retaining only those nodes $i$, $j$ and edges $e_{ij}$ for which $w_{ij}$ is 
	above the $p$-th percentile of all edge weights in the backboned network. 
	The network displayed here is constructed by setting $k = 20$ and $p = 50$, 
	where size of the node indicates normalized page rank.
	Topics are associated with distinct communities, found using the modularity algorithm of Clauset
	\textit{et al.} \cite{clauset2004finding}.} 
	\label{fig:inferred-topic-network-cusp}
\end{figure*} 
  
\section{Document-free topic networks}

An important application of the DST is the partial recovery of context- or document-dependent information
from aggregated time series data.
In natural language processing, many models of human language are statistical in nature and require original documents 
from which to infer values of parameters and perform estimation \cite{blei2003latent,dou2011paralleltopics}.
However, such information can be both expensive to purchase and require a large amount of physical storage space.
For example, the tweet corpus from which the labMT rank dataset used throughout this paper was originally derived is not inexpensive
and requires approximately 55 TB of disk space for storage \footnote{The dataset is available for purchase from Twitter at 
\www{http://support.gnip.com/apis/firehose/overview.html}.
The on-disk memory statistic is the result of \texttt{du -h <dirname>
| tail -n 1} on the authors'
computing cluster and so may vary by machine or storage system}.
In contrast, the dataset used here is derived from the freely-available LabMT word set and is less than 400 MB in size. 
If topics of relatively comparable quality can be extracted from this smaller and less expensive dataset, the 
potential 
utility to the scientific community at large, 
could be high.

\begin{figure*}[!th]
\centering
	\includegraphics[width=\textwidth]{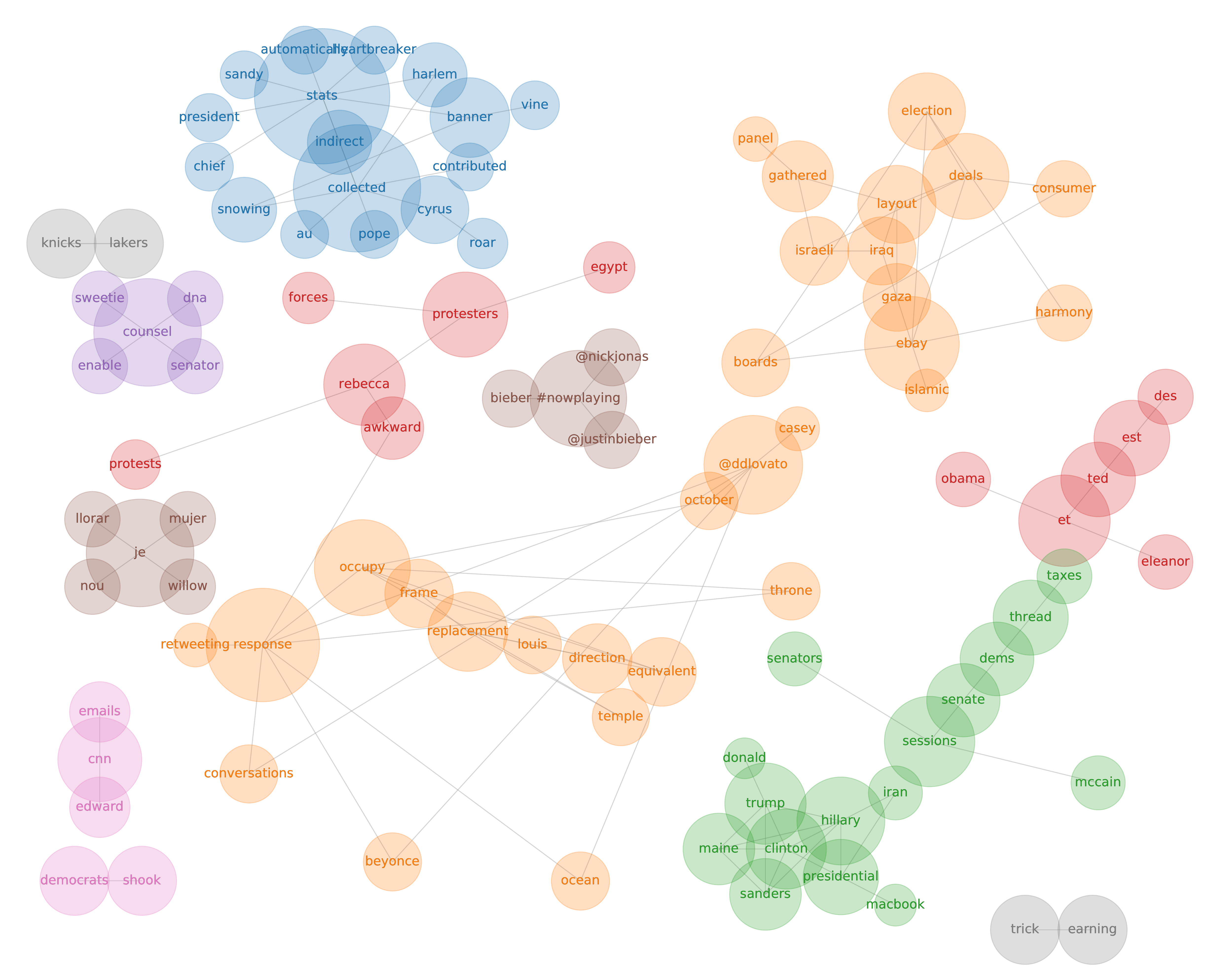}
	\caption{Topic network inferred from weighted spike indicator functions.
	At each point in time, words are ranked according to the value of their weighted spike
	indicator function and the top $k$ words are taken and linked pairwise for an 
	upper bound of $\binom{k}{2}$ additional edges in the network; if the edge between words $i$ and $j$ already exists, 
	the weight of the edge is incremented.
	The edge weight increment at time $t$ is given by $w_{ij,t} = \frac{R_{i,t} + R_{j,t}}{2}$,
	the average of the weighted spike indicator for words $i$ and $j$, 
	with the total edge weight thus given by $w_{ij} = \sum_t w_{ij,t}$.
	After initial construction, the backbone of the network is extracted using the method of Serrano \textit{et al.}
	\cite{serrano2009extracting}.
	The network is pruned further by retaining only those nodes $i$, $j$ and edges $e_{ij}$ for which $w_{ij}$ is 
	above the $p$-th percentile of all edge weights in the backboned network. 
	The network displayed here is constructed by setting $k = 20$ and $p = 50$, 
	where size of the node indicates normalized page rank.
	Topics are associated with distinct communities, found using the modularity algorithm of Clauset
	\textit{et al.} \cite{clauset2004finding}.}
	\label{fig:inferred-topic-network-shock}
\end{figure*}

We demonstrate that a reasonable topic model for Twitter during the time period of study 
can be inferred from the panel of rank time series alone.
This is accomplished via a multi-step meta-algorithm.
First, the weighted Shock Indicator Function $R_i$ is calculated for each word $i$.
At each point in time $t$, words are sorted by their respective shock indicator functions
as in Fig.\ \ref{fig:topk-shock}.
At time step $t$, the top $k$ words are taken and linked pairwise for an upper bound of $\binom{k}{2}$ 
additional edges in the network; if an edge already exists between word $i$ and $j$, it is incremented by the mean of the
words' respective weighted Shock Indicator Function $\frac{R_i + R_j}{2}$.
Performing this process for all time periods results in a weighted network of related words. 
The weights $w_{ij} = \sum_t \frac{R_{i,t} + R_{j,t}}{2}$ 
are large when the value of a word's weighted shock indicator
function is large or when a word is frequently in the top $k$, even if it is never near the top.
The resulting network can be large; to reduce its size, its backbone is extracted using the method of Serrano
\textit{et al.} \cite{serrano2009extracting} and further pruned by retaining only those nodes and edges for which the corresponding edge weights are at or above the $p$-th percentile 
of all weights in the backboned network.
Topics are associated with communities in the resulting pruned networks, found using the modularity algorithm of Clauset \textit{et al.} \cite{clauset2004finding}.

Fig.\ \ref{fig:inferred-topic-network-cusp} and Fig.\ \ref{fig:inferred-topic-network-shock} display the result of this procedure for $k = 20$ and $p = 50$.
Unique communities (topics) are indicated by node color. 
In the co-shock network (Fig.\ \ref{fig:inferred-topic-network-cusp}),
topics include, among others:
\begin{itemize}
\item Winter holidays and events 
(``valentines'', ``superbowl'', ``vday'',...);
\item U.S.\ presidential elections (``republicans'', ``barack'', ``clinton'', ``presidential'',...);
\item Events surrounding the 2016 U.S.\ presidential election in 
particular (``clinton's'', ``crooked'', ``giuliani'', ``jill'', ``stein'',...);
\end{itemize}
while the co-shock network displays topics pertaining to:
\begin{itemize}
\item popular culture and music (``bieber'', ``\#nowplaying'', ``@nickjonas'',
``@justinbieber'');
\item U.S. domestic politics 
(``clinton'', ``hillary'', ``trump'', ``sanders'', ``iran'', 
``sessions'',...);
\item and conflict in the Middle East (``gaza'', ``iraq'', ``israeli'', ``gathered'')
\end{itemize}
The predominance of U.S.\ politics at the exclusion of politics of other nations is likely because the labMT dataset contains predominantly English words.

\section{STAR and ADV comparison figures}
\label{app:star-vs-adv}

\begin{figure*}
\centering
\includegraphics[width=\textwidth]{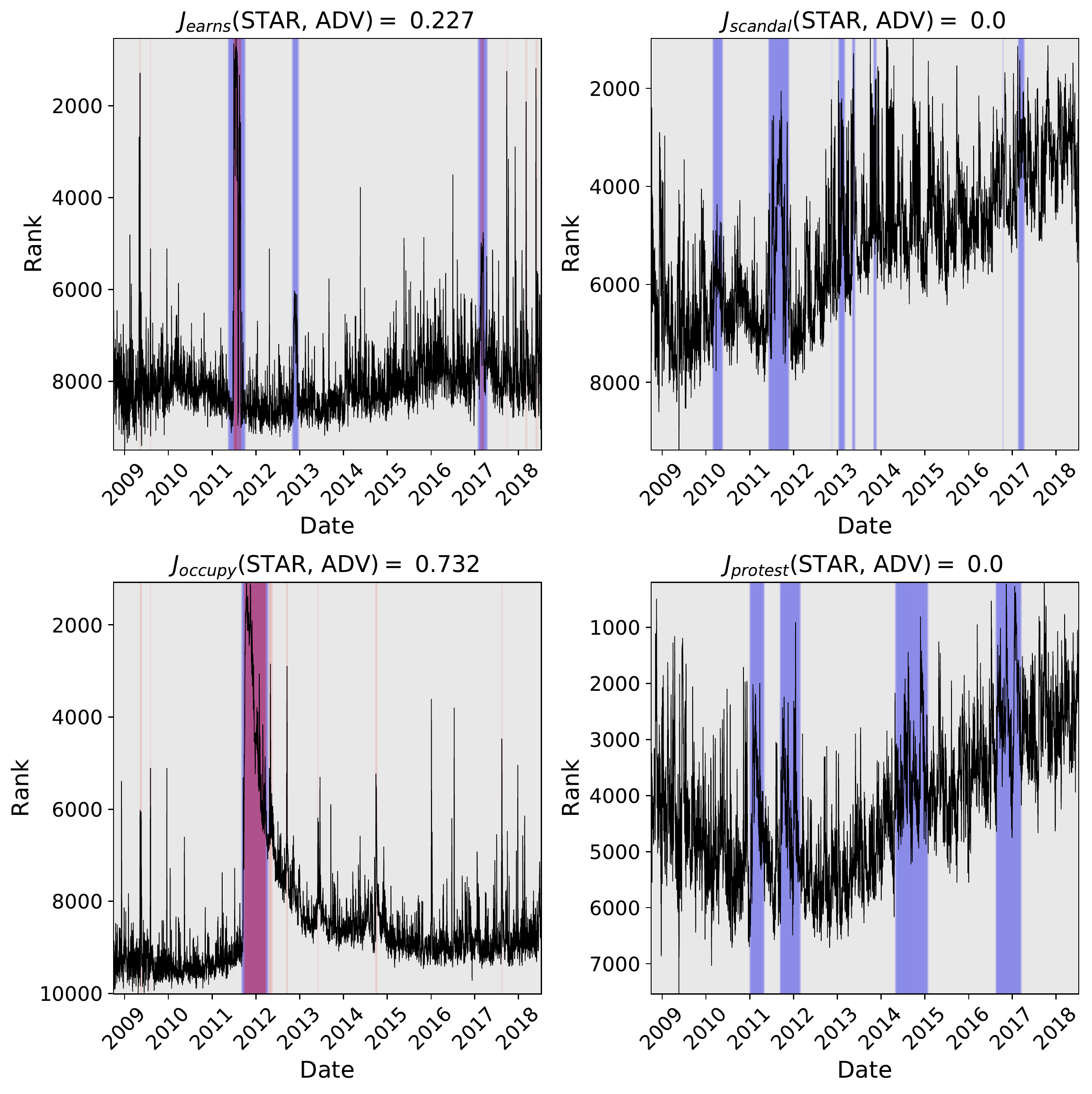}
\caption{
Comparison of STAR and ADV indicator windows for some words surrounding the ``Occupy Wall Street'' movement during 2010.
}
\label{fig:earns-scandal}
\end{figure*}

\begin{figure*}
\centering
\includegraphics[width=\textwidth]{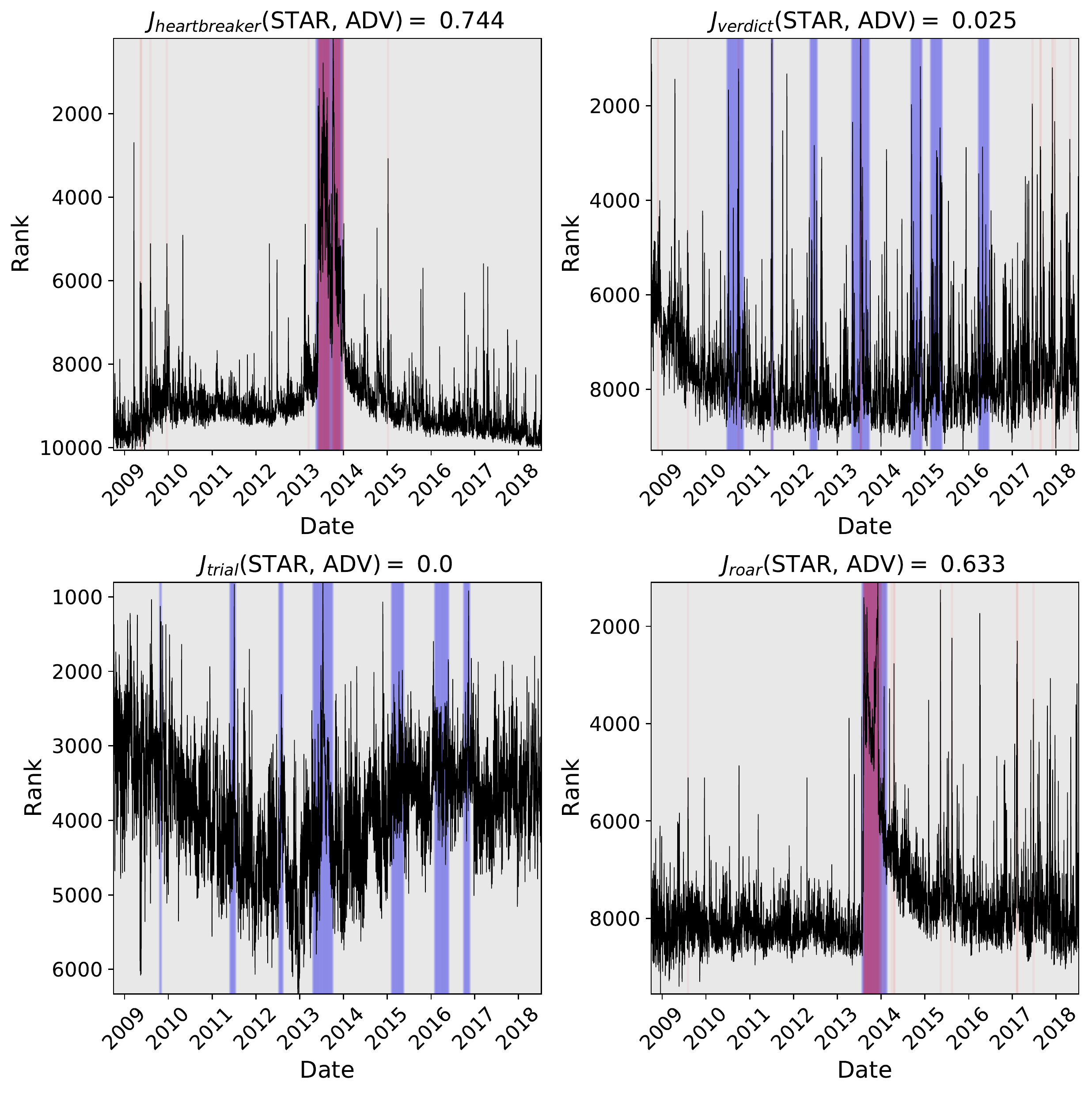}
\caption{
Comparison of STAR and ADV indicator windows for some words surrounding popular events (the release of a song called ``Heartbreaker'' by Justin Bieber and ``Roar'' by Katy Perry) and criminal justice-related events (the trial and acquittal of George Zimmerman). 
}
\label{fig:heartbreaker-verdict}
\end{figure*}

\begin{figure*}
\centering
\includegraphics[width=\textwidth]{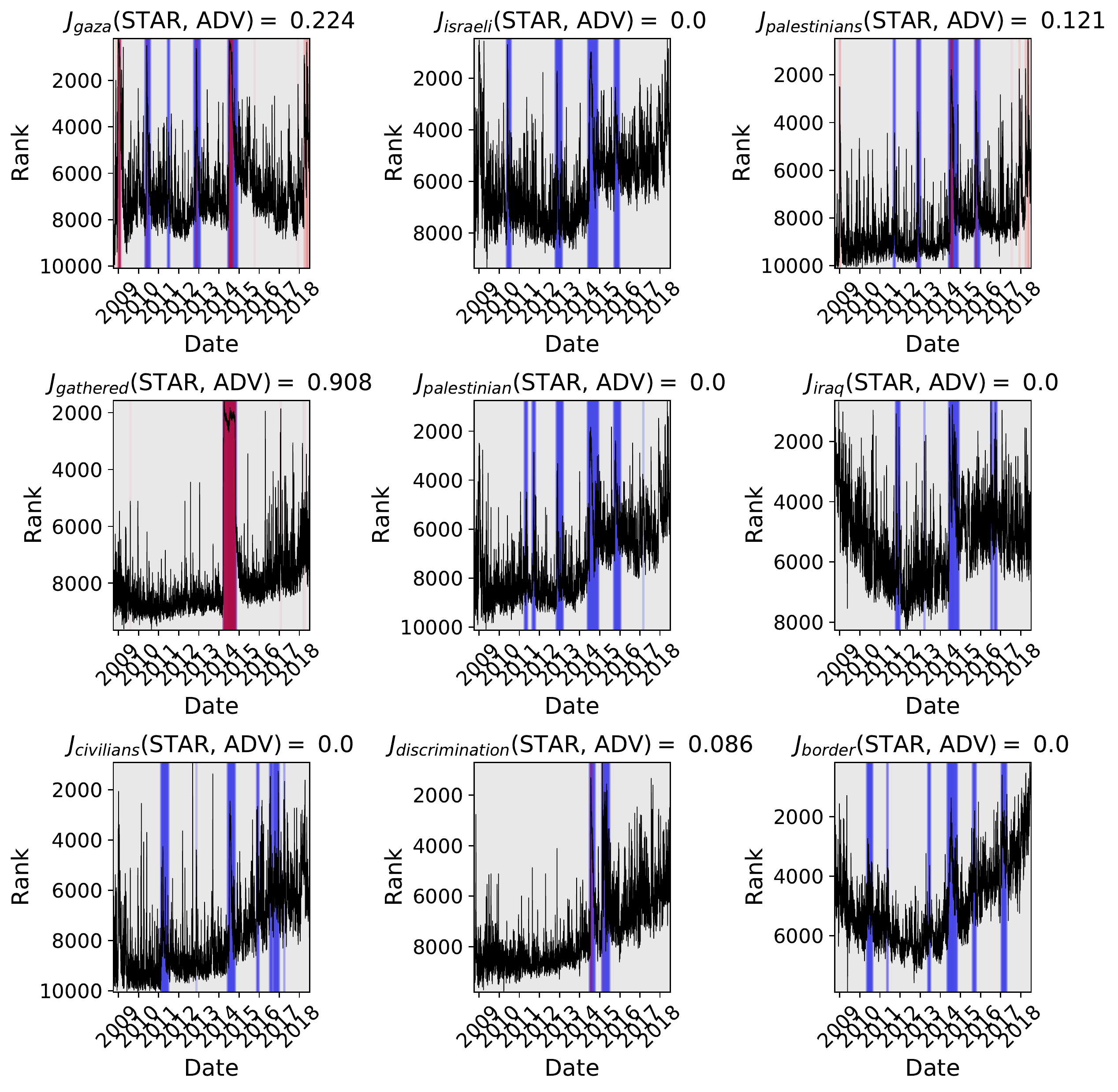}
\caption{
Comparison of STAR and ADV indicator windows for some words surrounding the Gaza conflict of 2014.
}
\label{fig:gaza-israeli}
\end{figure*}

\begin{figure*}
\centering
\includegraphics[width=\textwidth]{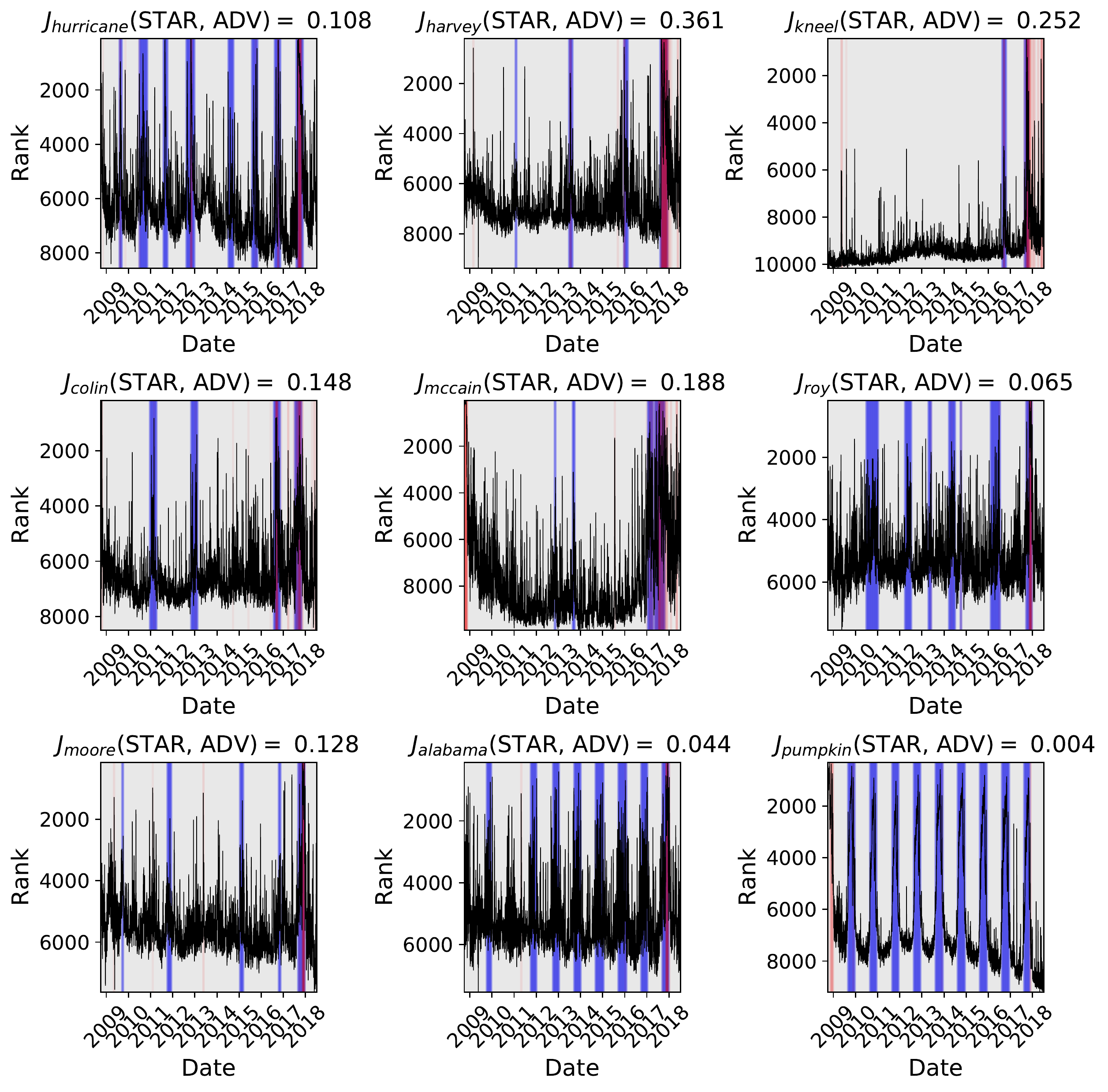}
\caption{
Comparison of STAR and ADV indicator windows for some words surrounding the autumn of 2017, including Hurricane Harvey, Colin Kaepernick's kneeling protests, John McCain, the electoral campaign of Roy Moore in the U.S.\ state of Alabama, and pumpkins (a traditional gourd symbolic of autumn in the U.S.)
}
\label{fig:hurricane-harvey}
\end{figure*}

\end{document}